\documentclass[preprint2]{aastex}
\usepackage[english]{babel}
\usepackage{adjustbox}
\usepackage{longtable}
\usepackage[utf8]{inputenc}
\usepackage{fixltx2e}
\usepackage{tabu}
\usepackage{placeins}
\usepackage{lscape}
\usepackage{amsmath}
\usepackage{graphicx}
\usepackage{float}
\usepackage{hyperref}

\shorttitle{The Peculiar ILOT SN 2010da} 
 \shortauthors{V. A. Villar et al.}

\begin{document}

\title{The Intermediate Luminosity Optical Transient SN 2010da: The Progenitor, Eruption and Aftermath of a Peculiar Supergiant High-mass X-ray Binary}

\author{V.~A.~Villar\altaffilmark{1,2}, E. Berger\altaffilmark{1}, R. Chornock\altaffilmark{3}, R. Margutti\altaffilmark{4}, T. Laskar\altaffilmark{5,6}, P. J. Brown\altaffilmark{7}, P. K. Blanchard\altaffilmark{1,2}, I. Czekala\altaffilmark{1}, R. Lunnan\altaffilmark{8}, M. T. Reynolds\altaffilmark{9}}
\altaffiltext{1}{Harvard Smithsonian Center for Astrophysics, 60 Garden St., Cambridge, MA 02138, USA;\href{mailto:vvillar@cfa.harvard.edu}{vvillar@cfa.harvard.edu}}
\altaffiltext{2}{NSF GRFP Fellow}
\altaffiltext{3}{Astrophysical Institute, Department of Physics and Astronomy, 251B Clippinger Lab, Ohio University, Athens, OH 45701,USA}
\altaffiltext{4}{New York University, Physics department, 4 Washington Place, New York, NY 10003, USA}
\altaffiltext{5}{Jansky Fellow, National Radio Astronomy Observatory}
\altaffiltext{6}{Department of Astronomy, University of California, Berkeley, CA 94720-3411, USA}
\altaffiltext{7}{George P. and Cynthia Woods Mitchell Institute for Fundamental Physics \& Astronomy, Texas A. \& M. University, Department of Physics and Astronomy, 4242 TAMU, College Station, TX 77843, USA}
\altaffiltext{8}{Department of Astronomy, California Institute of Technology, 1200 East California Boulevard, Pasadena, CA 91125, USA}
\altaffiltext{9}{Department of Astronomy, University of Michigan, 1085 S. University Ave, Ann Arbor, MI 48109-1107, USA}

\begin{abstract}
We present optical spectroscopy, ultraviolet to infrared imaging and X-ray observations of the intermediate luminosity optical transient (ILOT) SN 2010da in NGC\,300 ($d=1.86$ Mpc) spanning from $-6$ to $+6$ years relative to the time of outburst in 2010. Based on the light curve and multi-epoch SEDs of SN 2010da, we conclude that the progenitor of SN 2010da is a $\approx10-12$ M$_\odot$ yellow supergiant possibly transitioning into a blue loop phase. During outburst, SN 2010da had a peak absolute magnitude of M$_\mathrm{bol}\lesssim -10.4$ mag, dimmer than other ILOTs and supernova impostors. We detect multi-component hydrogen Balmer, Paschen, and Ca II emission lines in our high-resolution spectra, which indicate a dusty and complex circumstellar environment. Since the 2010 eruption, the star has brightened by a factor of $\approx 5$ and remains highly variable in the optical. Furthermore, we detect SN 2010da in archival \textit{Swift} and \textit{Chandra} observations as an ultraluminous X-ray source ($L_\mathrm{X}\approx6\times10^{39}$ erg s$^{-1}$). We additionally attribute He II 4686\AA\ and coronal Fe emission lines in addition to a steady X-ray luminosity of $\approx 10^{37}$ erg s$^{-1}$ to the presence of a compact companion. 
\end{abstract}


\section{Introduction}

Between the luminosities of the brightest novae $(\mathrm{M_V}\approx -10$; \citealt{hachisu2014light}) and the dimmest supernovae $(\mathrm{M_V}\approx -14$; \citealt{zampieri2003peculiar}), there is a dearth of well-studied optical transients (see \citealt{kasliwal2012}). In the last decade, we have begun to fill in this gap with a number of exotic events such as luminous red novae \citep{kulkarni2007unusually}, luminous blue variable (LBV) outbursts and other ``supernova impostors'' (e.g. \citealt{van2000sn,pastorello2007giant,berger2009,tartaglia2015}). Additionally, there are expected events which have not been definitively observed, such as ``failed'' supernovae \citep{kochanek2008survey}. Following \cite{berger2009}, we will collectively refer to these events as intermediate luminosity optical transients (ILOTs).

The link between ILOTs and their progenitors remains elusive, especially for ILOTs surrounded by dense circumstellar media (CSM). Brighter dusty ILOTs, such as the great eruption of Eta Carinae \citep{davidson1997eta} or SN 1954J \citep{van2005supernova}, are attributed to LBV outbursts; however, the progenitors of dimmer events are under debate with a larger pool of viable origins. For example, theorized progenitors of the famous dusty ILOTs, such as NGC\,300 OT2008-1 and SN 2008S have ranged from mass loss events of yellow hypergiants \citep{berger2009}, to mass transfer from an extreme AGB star to a main sequence companion \citep{kashi2010ngc}, to low luminosity electron-capture supernovae \citep{thompson2009new,2015arXiv151107393A}. Each of these interpretations shares the common theme of marking an important point in the evolution of relatively massive stars ($\gtrsim9 M_\odot$). 

Adding to the diversity of ILOTs is the possibility of optical transients within X-ray binary systems. High mass X-ray binaries (HMXBs) consist of a massive star and a compact object (e.g. a neutron star or a black hole) and produce X-rays as material accretes onto the compact object through a variety of channels \citep{lewin1997x,reig2011x}. A relatively new subclass of HMXBs known as obscured HMXBs are cloaked in a high density of local material ($N_\mathrm{H}\sim10^{23}-10^{24}$ cm$^{-2}$; \citealt{chaty2007optical,tomsick2009xmm}). While the primary stars of these systems are largely unknown, several have been shown to be supergiants exhibiting B[e] phenomena \citep{clark1999near,chaty2004revealing,kaplan2006long}. These systems likely produce their dense circumstellar material through either a continuous wind or periodic outbursts which have not yet been observed.

In this work we report data from a five-year, multiwavelength (X-ray, ultraviolet, optical and infrared) observational campaign of the dusty ILOT SN 2010da which was discovered in the nearby galaxy NGC\,300 \citep{monard2010}. We show that SN 2010da exhibits many features shared amongst dusty ILOTs, such as striking Balmer emission and optical variability on the order of months, but it is the only ILOT to sit in an intermediate range between extremely dusty red transients such as SN 2008S and the bluer, brighter LBV outbursts. Additionally, SN 2010da is the first ILOT to be a member of a high mass X-ray binary which undergoes an ultraluminous X-ray outburst ($\sim$ $10^{40}$ erg s$^{-1}$). Previous work on SN 2010da (\citealt{binder2011,binder2016}) concluded that the progenitor is a massive ($\gtrsim 25$ M$_\odot$) luminous blue variable using limited \textit{HST} photometry. However, from our broadband photometry and spectroscopy we infer that SN 2010da originated from an intermediate mass ($\sim 10-12\mathrm{M}_\odot$), variable yellow supergiant progenitor which is now transitioning into a blue loop phase of its evolution. We discuss these conflicting interpretations and the importance of comprehensive, multi-wavelength coverage of ILOTs.

\section{Observations}
\label{sec:observations}

SN 2010da was discovered in NGC\,300 on 2010 May 23.169 UT by \cite{monard2010} with an unfiltered magnitude of $16.0\pm0.2$, corresponding to $M\approx-10.3$ assuming a distance of 1.86 Mpc \citep{rizzi2006} and a foreground extinction of E(B-V)$=0.011$ \citep{schlafly2011measuring}. We neglect addition extinction from NGC\,300 based on our observed \textit{Swift} colors (Section \ref{sec:anal:outburst}). Throughout this paper, Epoch 0 will refer to the discovery date, 2010 May 23. Prior to discovery, NGC\,300 was behind the Sun, although \cite{monard2010} reported an upper limit of $\lesssim15.5$ mag on May 6. Archival \textit{Spitzer} data indicated that the source began brightening in the infrared at least 150 days before the optical discovery \citep{atel2648}. Multi-wavelength follow-up, spanning from the radio to X-rays, revealed that despite its supernova designation, SN 2010da was likely an outburst of a massive star enshrouded by dust \citep{elias2010sn,chornock2010spectroscopy,atel2660}. This conclusion was reaffirmed by archival \textit{Spitzer}/IRAC observations of the dusty progenitor \citep{khan2010mid,atel2638}, but the lack of extinction in the spectral energy distribution (SED) suggested that some dust had been destroyed during the outburst \citep{brown2010swift,atel2640}.  Early spectroscopic followup revealed narrow emission features (FWHM $\approx 1000$ km s$^{-1}$) with no signs of P-Cygni profiles \citep{elias2010sn}. Hydrogen Balmer, Fe II and He I emission lines provided further support for interaction with a dense CSM surrounding the progenitor.

The transient was also detected in the X-rays and UV with the \textit{Swift} X-ray Telescope (XRT) and Ultraviolet/Optical Telescope (UVOT), respectively \citep{atel2639,brown2010swift}. Additionally, 3$\sigma$ upper limits of F$_\nu\lesssim$  87 (4.9 GHz), $\lesssim$ 75 (8.5 GHz), and $\lesssim$ 225 (22.5 GHz) $\mu$Jy were obtained with the Karl G. Jansky Very Large Array \citep{atel2658}. Following the event, we monitored SN 2010da in the near-infrared (NIR) and optical using Gemini and Magellan. We report below our ground-based imaging and spectroscopy, as well as an analysis of archival \textit{Spitzer, Hubble, Swift} and \textit{Chandra} observations.

\subsection{\textit{Spitzer} Infrared Imaging}\label{sec:spitzer_reduction}
\begin{figure*}[h!]
\centering
\includegraphics[width=\textwidth]{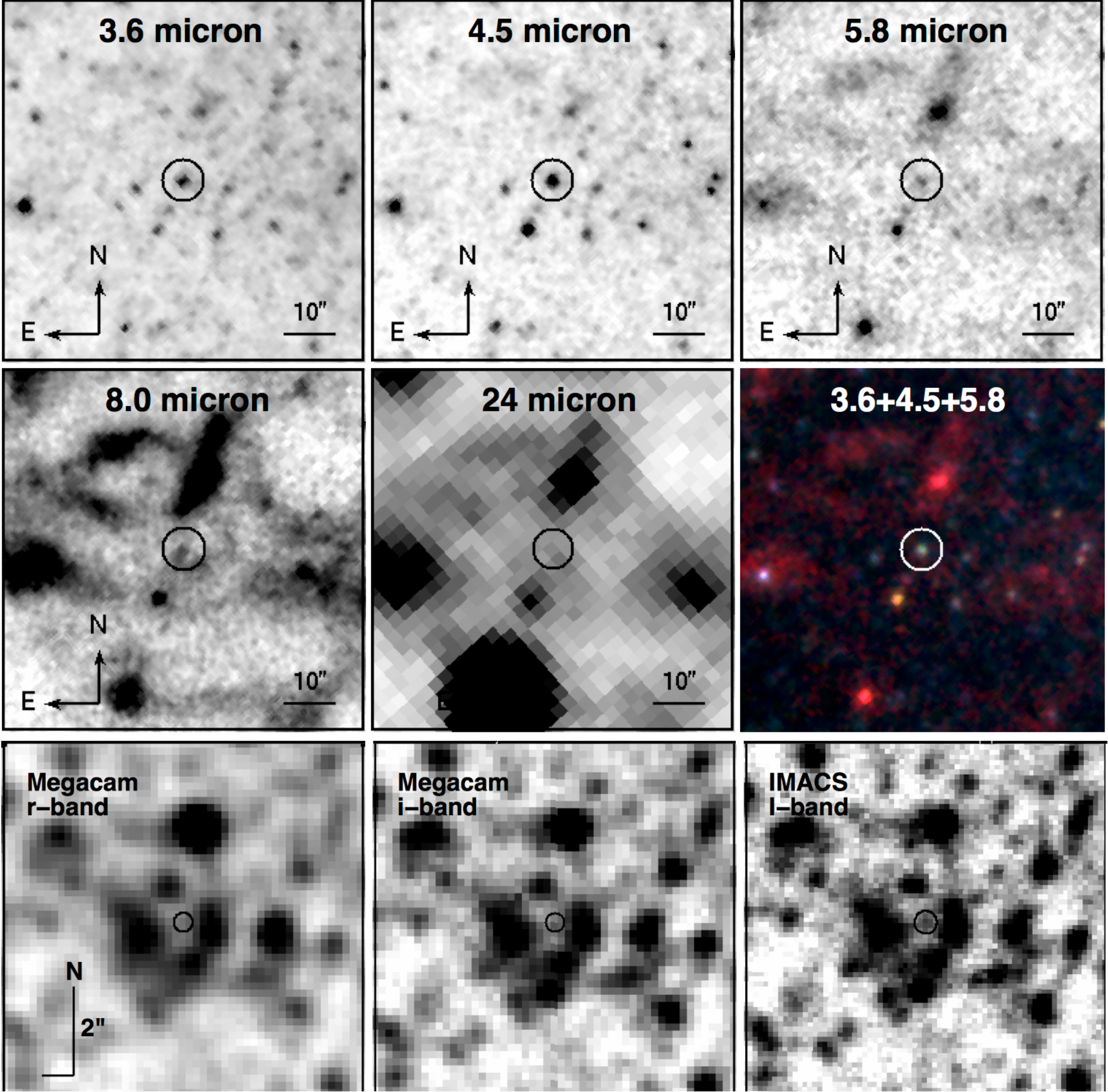}
\caption{Top rows: \textit{Spitzer} images of the SN 2010da progenitor. The right panel of the middle row shows a false color image combining the 3.6 (blue), 4.5 (green) and 5.8 (red) $\mu$m images. Bottom row: Archival MegaCam and IMACS images. The progenitor is only detected in the IMACS I-band image.}
\label{fig:spitzer_progen}
\end{figure*}

We obtained publicly available \textit{Spitzer} images spanning from 2003 November 21 to 2016 March 19 (see Table \ref{tab:spitz} for program IDs; \citealt{lau2016rising}). This data set extends several years before and after the event, but no data are available within a four month window surrounding the optical discovery. We used data from the InfraRed Array Camera (IRAC) in the 3.6 and 4.5 $\mu$m bands through both the original and ``warm" \textit{Spitzer} missions, and we use IRAC data in the 5.8 and 8.0 $\mu$m bands available prior to the 2010 eruption. Additionally, we used photometry from the Multiband Imagine Photometer (MIPS) in the 24 $\mu$m band prior to the discovery of the transient. We processed the \textit{Spitzer} data with the {\tt Mopex} package, which creates a mosaic of the dithered \textit{Spitzer} images. For the IRAC images, we used a drizzling parameter of 0.7 and an output pixel scale of $0.4''$. For the MIPS images, we used the same drizzling parameter but with an output pixel scale of $1.8''$. Images of the field in the \textit{Spitzer} bands are shown in Figure \ref{fig:spitzer_progen}.

\begin{figure*}
\centering
\includegraphics[width=\textwidth]{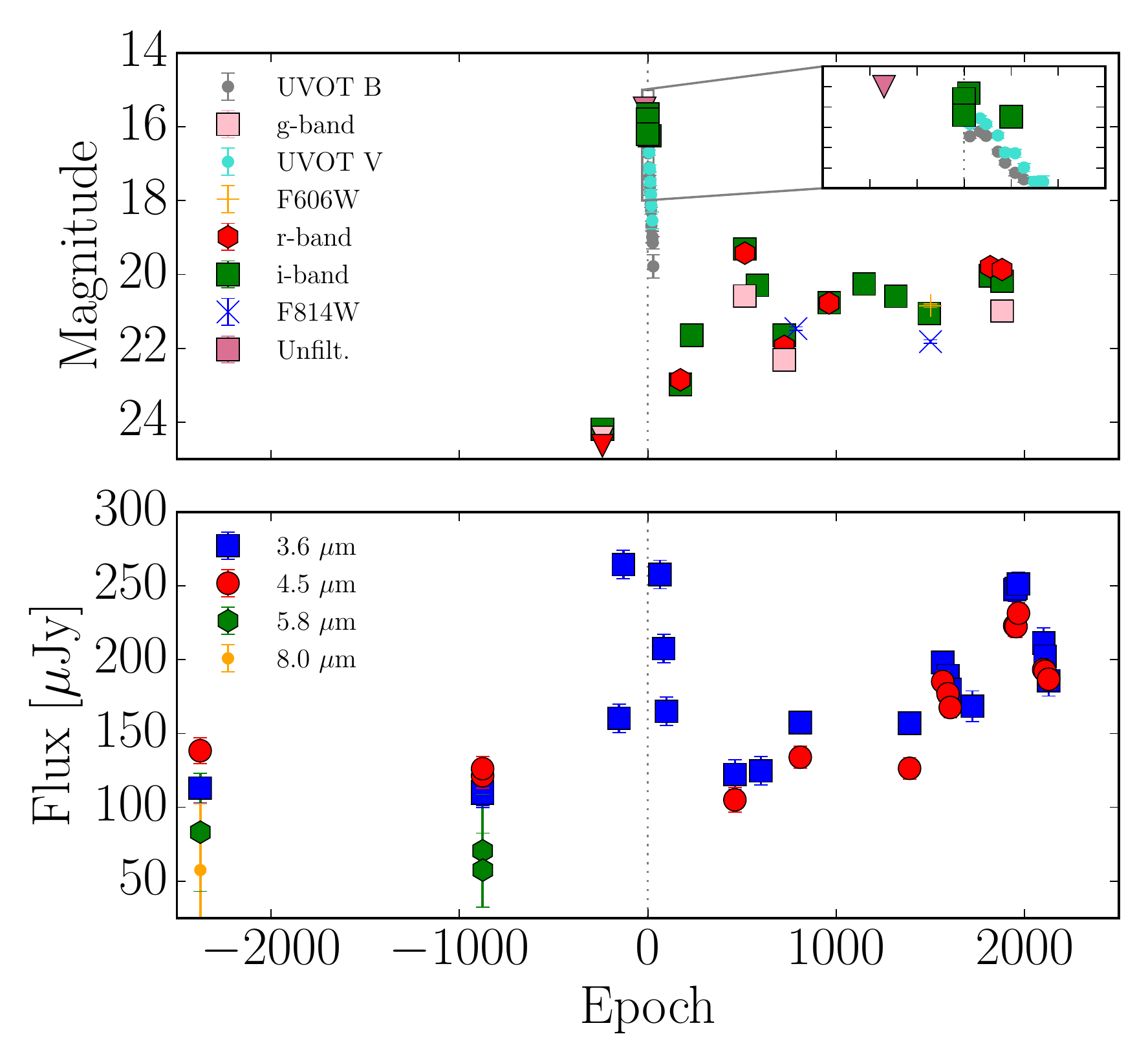}
\caption{The \textit{Spitzer}, ground-based optical, \textit{HST} and \textit{UVOT} light curves of SN 2010da, spanning 6 years before to 6 years after the 2010 eruption. Epoch 0 refers to 2010 May 23.169 UT, the date of discovery. Downward facing triangles are 3$\sigma$ upper limits.}
\label{fig:lc_opt}
\end{figure*}

We performed aperture photometry using DS9's {\tt Funtools}. For the IRAC photometry, we used an aperture of 3 native IRAC pixels (corresponding to 3.66'') with an inner and outer background annulus radii of 3 (3.66'') and 7 (8.54'') native pixels, respectively. These radii have calculated aperture correction factors for point sources in the IRAC Instrument Handbook. For the MIPS 24 $\mu$m photometry we used an aperture size of 3.5'' with no background annulus, following the MIPS Instrument Handbook. We calculated the flux uncertainties following Equation 1 in \cite{laskar2011}. The observations are summarized in Table \ref{tab:spitz}, and the \textit{Spitzer}/IRAC light curves at 3.6 and 4.5 $\mu$m are shown in Figure \ref{fig:lc_opt}. Our photometric results are consistent with those presented in \cite{lau2016rising}.

\subsection{Ground-based Near-Infrared Imaging}
We obtained near-infrared imaging observations with the FourStar Infrared Camera \citep{persson2013fourstar} on the Magellan/Baade 6.5m telescope at the Las Campanas Observatory in Chile on three epochs: 2011 December 7 ($J$, $H$, $K_s$), 2015 July 31 ($H$, $K_s$) and 2015 August 18 ($J$, $H$, $K_s$). We calibrated, aligned and co-added each of these observations using the {\tt FSRED} package\footnote{http://instrumentation.obs.carnegiescience.edu/FourStar/\\SOFTWARE/reduction.html}. Each image was calibrated using the 2MASS Point Source Catalog, and the magnitude of the transient was measured using aperture photometry. The results are summarized in Table \ref{tab:fourstar}.

\subsection{Ground-based Optical Imaging}
We obtained optical imaging observations with the Low Dispersion Survey Spectrograph 3 (LDSS-3, upgraded from LDSS-2 \citealt{allington1994low}) and the Inamori-Magellan Areal Camera \& Spectrograph (IMACS; \citealt{dressler2006imacs}) on the Magellan Clay and Baade 6.5m telescopes at the Las Campanas Observatory, respectively, in the \textit{gri} filters spanning from $\approx610$ days before to $\approx1900$ days after the optical discovery. In our earliest IMACS I-band image (at Epoch $-609$), we detect the object with $24.2\pm0.2$ mag (see Figure \ref{fig:spitzer_progen}). However, we do not detect a source at the location of SN 2010da in pre-transient \textit{gri} images taken with the Magellan/Clay wide field imager MegaCam (at Epoch $-183$; \citealt{mcleod2015megacam}). We use the MegaCam images in each filter as templates for image subtraction. For all other ground-based optical imaging observations, we performed image subtraction using the {\tt ISIS} package \citep{alard2000image}. We then performed aperture photometry on the subtracted images and calibrated to southern standard stars listed in \cite{standardsouthern}. The photometry is summarized in Table \ref{tab:ldss}.

\subsection{\textit{HST} Optical Imaging}
SN 2010da was observed by the \textit{Hubble Space Telescope} Advanced Camera for Surveys (ACS) on 2012 July 18 (Program 12450) and 2014 July 9 (Program 13515). The object was observed in the F814W filter in both programs (2224 s and 2548 s exposure times, respectively) and in the F606W filter with program 13515 (2400 s). We processed the data using the standard {\tt PyDrizzle} pipeline in {\tt PyRAF} which supplies geometric distortion corrections to combine undersampled, dithered images from \textit{HST}. We scaled the pixel size by 0.8 for a final pixel scale of $0.032''$. We detected a source coincident with the position of SN 2010da, and using five objects detected in the field from the 2MASS Point Source Catalog, we determined a position of $\alpha=$ 00$^h$55$^m$04.86$^s$, $\delta= -37^o$41$'$43.8$''$ (J2000) with $0.3''$ uncertainty in both coordinates. This is in good agreement (within $1\sigma$) with previous results \citep{monard2010,binder2011}. With the high resolution of \textit{HST}, SN 2010da appears isolated, and we used aperture photometry to measure its magnitude. These magnitudes are listed in Table \ref{tab:hst} and are in good agreement with those reported by \cite{binder2016}.

\subsection{Optical Spectroscopy}
We obtained medium- and high-resolution spectra of SN 2010da using: the Gemini South Multi-Object Spectrograph (GMOS;\citealt{davies1997gmos}) located in the southern Gemini Observatory in Chile; IMACS,  the Magellan Inamori Kyocera Echelle (MIKE; \citealt{bernstein2003mike}) spectrograph on the 6.5m Magellan/Clay telescope; and the Magellan Echellette Spectrograph (MagE; \citealt{marshall2008mage}) also mounted on the Magellan/Clay telescope. Table \ref{tab:spec} summarizes these observations. We used standard {\tt IRAF} routines to process the spectra and applied wavelength calibrations using HeNeAr arc lamps. The MIKE spectra were processed using a custom pipeline and calibrated using ThAr arc lamps. We used our own IDL routines to apply flux calibrations from observations of standard stars (archival in the case of Gemini) and correct for telluric absorption. We estimate the resolution of each spectrum (see Table \ref{tab:spec}) using their associated arc lamp spectra. All spectra are corrected for air-to-vacuum and heliocentric shifts.

\subsection{\textit{Swift}/UVOT Imaging}
The Swift/UVOT data was processed using the method of the Swift Optical/Ultraviolet Supernova Archive (SOUSA; \citealt{Brown_etal_2014_SOUSA}).  We combined pre-outburst observations from December 2006 and January 2007 into templates from which the underlying host galaxy count rate was measured.  A 3$''$ aperture was used with aperture corrections based on an average PSF.  A time-dependent sensitivity correction was used (updated in 2015) and AB zeropoints from \citet{Breeveld_etal_2011}.  The photometry is summarized in Table \ref{tab:uvotphot}.

\subsection{X-ray Spectral Imaging}
\label{sec:XrayObs}
We aggregated archival X-ray observations from the \textit{Swift}/XRT, the \textit{Chandra} X-ray Observatory and \textit{XMM-Newton}. These X-ray observations span from 2000 December 26 to 2014 November 17, including the outburst period. The source was undetected with \textit{XMM-Newton}, and we use the $3\sigma$ upper limits obtained by \cite{binder2011}.

The XRT observations were made before, during and after the 2010 outburst, and an X-ray source coincident with SN 2010da is detected in all three regimes. These observations are publicly available from the \textit{Swift} Archive \citep{evans2009methods}, and the XRT photometry and spectra are automatically generated through this database.

We used three archival \textit{Chandra} observations from 2010 September 24 (Obs. ID: 12238; PI: Williams), 2014 June 16 (Obs. ID: 16028; PI: Binder) and 2014 November 17 (Obs. ID: 16029; PI: Binder). All observations were made using the Advanced CCD Imaging Spectrometer (ACIS-I) with similar exposure times (63.0 ks, 64.24 ks and 61.27 ks, respectively). We analyzed the observations using {\tt CIAO} version 4.7 and {\tt CALDB} version 4.6.7 using standard extraction procedures. We performed photometry with {\tt WAVDETECT} using an annular background region with an inner radius of $24.6''$ and a width of $4.9''$ centered on the source. The results are summarized in Table \ref{tab:chandra}. We extracted spectra of the source using the built-in function {\tt specextract}.

\FloatBarrier

\section{The Multi-wavelength Properties of SN 2010da, its Progenitor, and its Progeny}

\subsection{Light Curve and Spectral Evolution}
\subsubsection{The Progenitor}\label{sec:anal:lcevolve:progen}
We are able to constrain the progenitor properties using the \textit{Spitzer} (3.6, 4.5, 5.8, 8 and 24 $\mu$m) and MegaCam/IMACS (\textit{gri}) observations. We note that the location of SN 2010da was observed in the $i'$-band on both 2008 September 09 and 2009 November 25 by IMACS and MegaCam, respectively. The MegaCam/IMACS observations are summarized in Table \ref{tab:ldss}. The progenitor IMACS detection and MegaCam upper limit are consistent with a magnitude $\sim 24.2$. The $gr$ upper limits were both obtained with MegaCam on 2009 November 25. The location of SN 2010da was observed five times by \textit{Spitzer} before the transient, ranging between 2003 November 21 and 2010 January 14. These observations are summarized in Table \ref{tab:spitz} We find no significant change in the color and brightness between the pre-eruption observations.

\begin{figure*}[hp]
\begin{center}
\includegraphics[width=\textwidth]{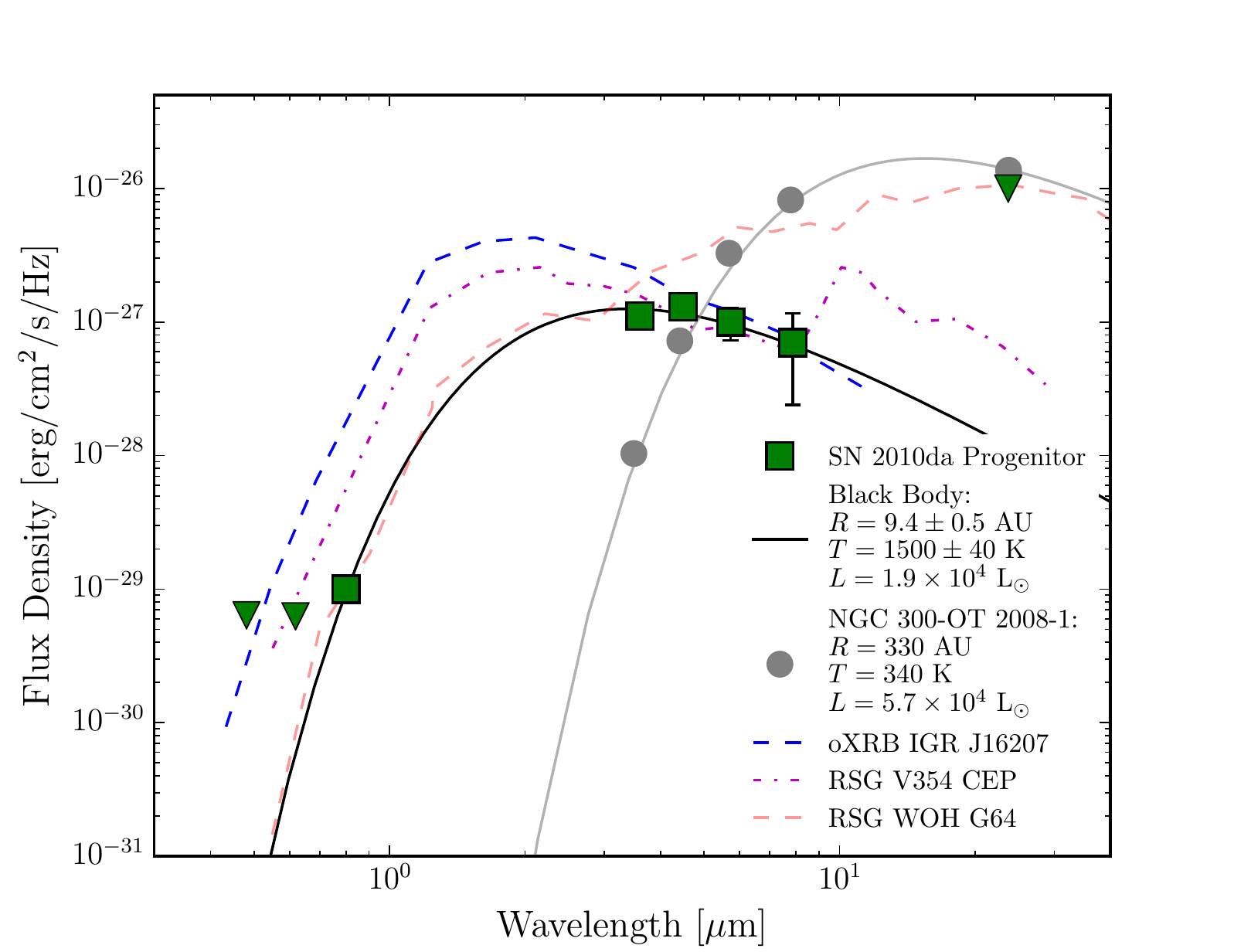}
\caption{Spectral energy distribution of the progenitor of SN 2010da (green squares) with a blackbody fit (black). Also shown are an obscured HMXB (blue dashed line, IGR J16207-5129;  \citealt{tomsick2006identifications}) and two RSGs (purple and pink lines, see Section \ref{sec:anal:lcevolve:progen} for a discussion of these objects; \citealt{van1999mass, mauron2011mass}). The latter three objects have been arbitrarily scaled to match the luminosity of the progenitor of SN 2010da. NGC\,300's other well-known impostor, NGC\,300 OT2008-1, as well as its best-fit blackbody, are shown in grey for comparison. Although NGC\,300 OT2008-1 and SN 2010da are spectroscopically similar, the progenitor of SN 2010da is obscured by significantly warmer dust. Downward facing triangles are 3$\sigma$ upper limits.} 
\label{fig:progen_sed}
\end{center}
\end{figure*}

To create a progenitor SED, we average the two pre-eruption \textit{Spitzer} observations in the 3.6 and 4.5 $\mu$m filters and compiled the other detections. The SED of the progenitor is well fit by an unabsorbed blackbody spectrum with $T =1500\pm40$ K and $R =9.4\pm0.5$ AU ($\chi^2_r=1.2$ for d.f. = 3). These parameters correspond to a bolometric luminosity of $L=(1.92\pm0.26)\times10^4$ L$_\odot$, suggesting a $\sim 15$ M$_\odot$ main sequence progenitor if we assume solar metallicity \citep{meynet2000}. The large radius and cool temperature of this fit imply that the progenitor is surrounded by dust. The progenitor SED is shown in Figure \ref{fig:progen_sed} along with several red supergiants (RSGs) and the progenitor of a previous ILOT in NGC\,300 (NGC\,300 OT2008-1; \citealt{berger2009}).  Also shown is the SED of an obscured HMXB (IGR J16207-5129; \citealt{tomsick2006identifications}). The progenitor SED peaks between the typically bluer obscured HMXBs and the redder ILOTs such as NGC\,300 OT2008-1. The SEDs of RSGs seem to bridge this gap, owing their SED variability to diverse geometries (e.g. WOH G64 has notable IR excess possibly due to a dusty torus along the line of sight; \citealt{ohnaka2008spatially}), although neither RSG fits the observed SED.

\subsubsection{The 2010 Outburst}\label{sec:anal:outburst}

SN 2010da was discovered at its brightest known magnitude of $m_{\mathrm{unfiltered}}=16.0\pm0.2$. It is unclear if SN 2010da was caught at its true peak brightness, but the optical 15.5 mag upper limit 18 days prior and a slight rise in the \textit{Swift}/UVOT light curve hints that SN 2010da was discovered near its peak luminosity. An increase in IR flux is seen in the IRAC data as early as $\approx 130$ days before the optical discovery.  The full rise and fall caught by \textit{Spitzer} spans $\approx250$ days, as shown in Figure \ref{fig:lc_opt}.

\begin{figure*}[ph]
\centering
\includegraphics[width=\textwidth]{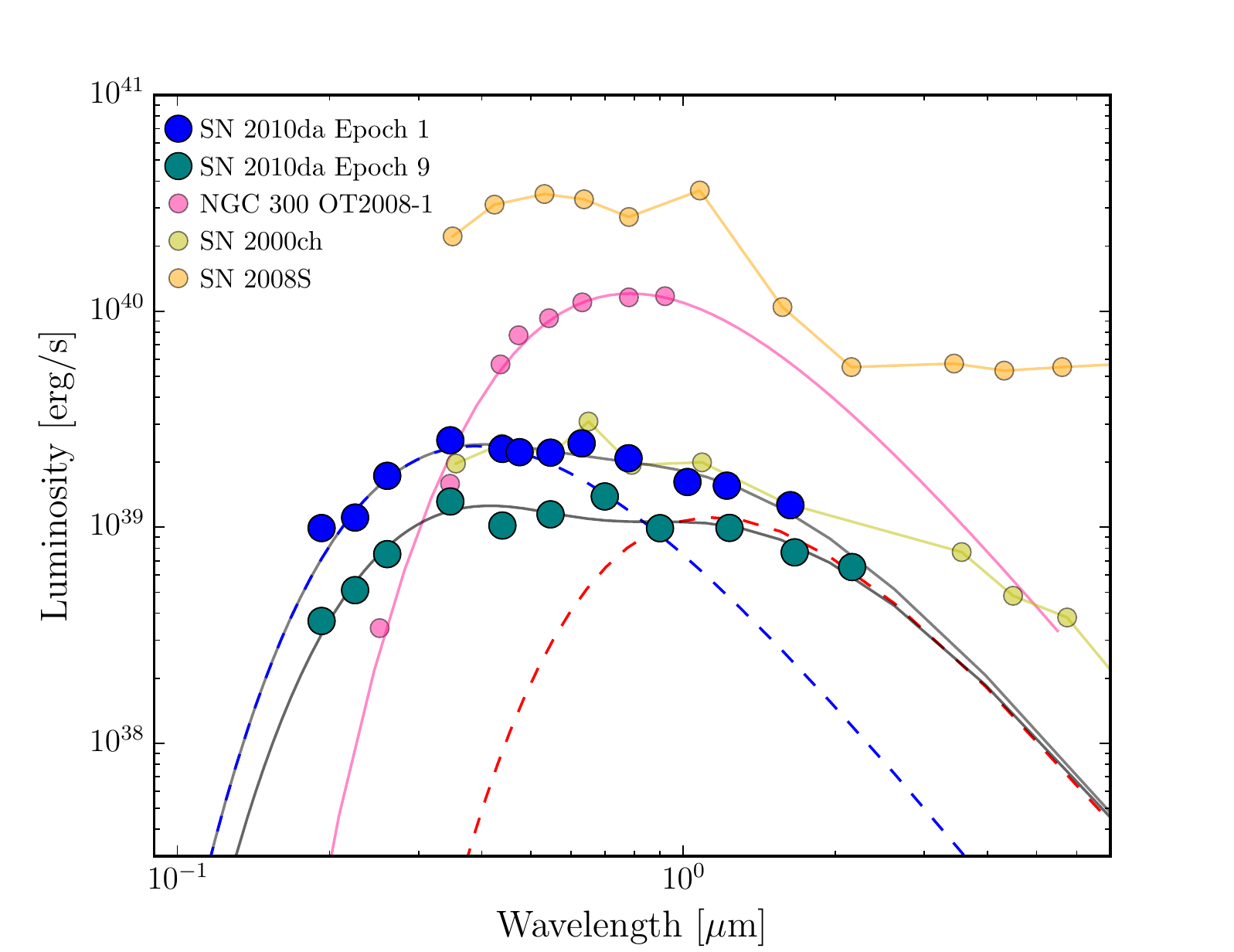}
\caption{The spectral energy distribution of SN 2010da 1 day (dark blue points) and 9 days (dark cyan points) after discovery. Both SEDs are fit with a two-component blackbody model. The total fit is shown in black, while the components of the outburst SED model are shown in blue and red for the first epoch. Also shown are the SEDs of a three similar ILOTs during outburst, NGC\,300 OT2008-1 (pink), SN 2000ch (yellow) and SN 2008S (orange). NGC\,300 OT2008-1 and SN 2008S are ILOTs with very red progenitors, while SN 2000ch is an LBV-like star.}
\label{fig:sed_explosion}
\end{figure*}

During the 2010 outburst, the SED of SN 2010da is well fit by two unabsorbed blackbodies at $\sim 0.2 - 1.7$ $\mu$m: a hotter blackbody with $T_{H,1}=9440\pm280$ K and $R_{H,1}=1.59\pm0.14$ AU and a cooler blackbody with $T_{C,1}=3230\pm490$ K and $R_{C,1}=9.5\pm2.9$ AU ($\chi^2_r\approx0.8$ for d.f. $= 6$), as shown in Figure \ref{fig:sed_explosion}. These black bodies have a combined bolometric luminosity of $L=(1.3\pm0.4)\times10^6$ L$_\odot$, about 60 times more luminous than the progenitor. Nine days later, the SED is consistent with similar blackbodies, although the larger one has cooled ($T_{C,2}=2760\pm250$ K, $R_{C,2}=10.5\pm1.6$ AU), while the hotter one remains at roughly the same temperature ($T_{H,2}=9080\pm330$ K, $R_{H,2}=1.25\pm0.13$ AU). The radius of the cooler, larger blackbody component is consistent with the estimated pre-eruption radius  ($R_{C,2}\approx R_{C,1}\approx R_{C,0}$) but has a temperature that is twice as high ($T_{C,2}\approx 2T_{C,0}$). These relations are summarized in Table \ref{tab:BB_fits}. This indicates that at least some dust in the original shell survived the outburst and has heated up.

\begin{figure}[ht!]
\centering
\includegraphics[width=0.45\textwidth]{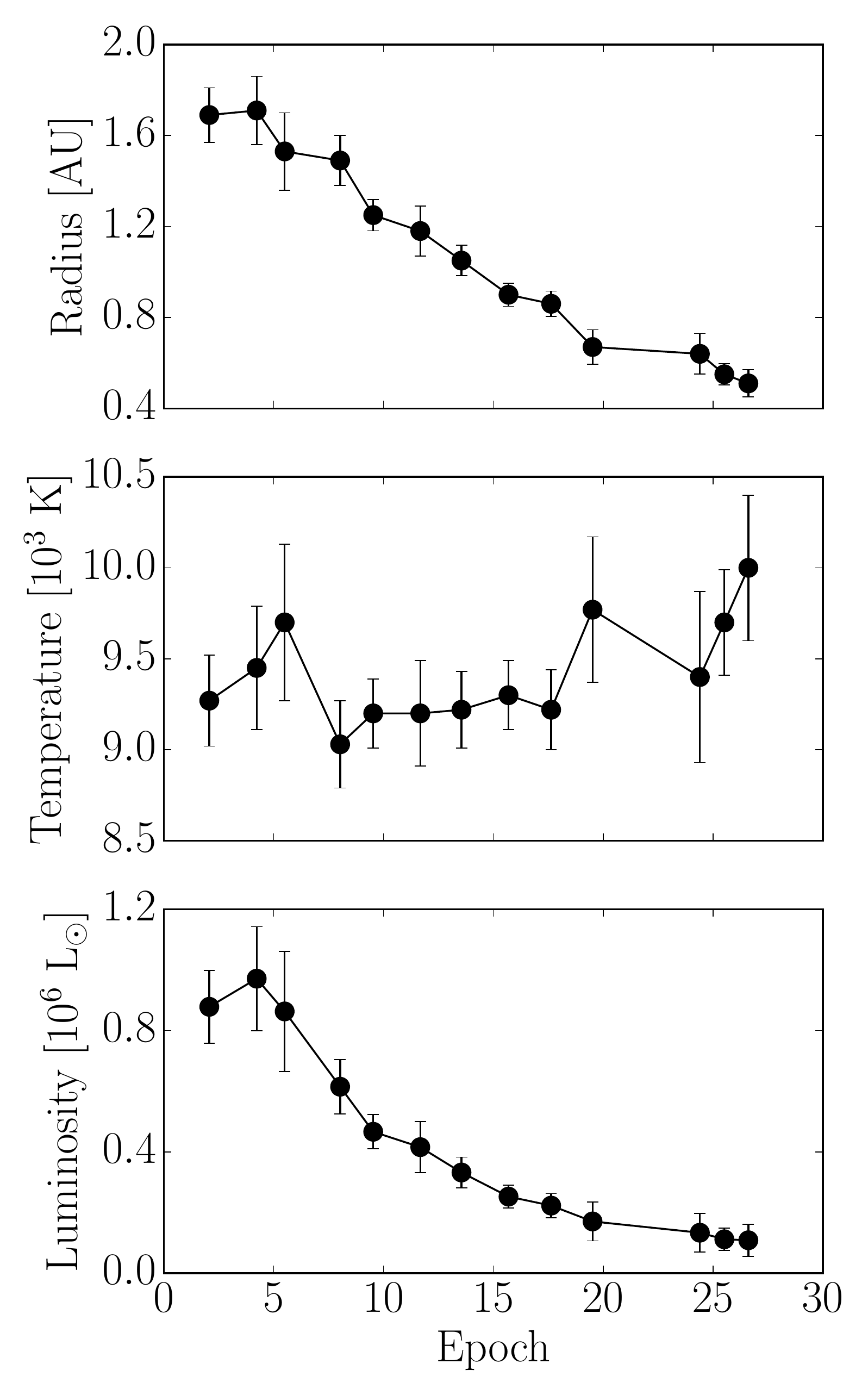}
\caption{We fit the \textit{UVOT} dataset to a blackbody as a function of time. We find that the blackbody radius recedes (top), while the temperature remains roughly constant (middle). The estimated luminosity decreases with the receding radius (bottom)}
\label{fig:bb_rad_temp}
\end{figure}

The UVOT data trace the evolution of the hotter blackbody detected in the initial outburst. The blackbody radius decreases from about 1.7 to 0.55 AU over the ten days of observations while remaining at a steady temperature of $\approx 9200$ K, as shown in Figure \ref{fig:bb_rad_temp}. This is consistent with a receding photosphere of the initial outburst. We can use this observation to constrain the radius of the progenitor/surviving star to $\lesssim 0.55$ AU, since we expect the photosphere of the eruption to exceed the radius of the star at all times.

\subsubsection{The Progeny}
About 200 days after its discovery, SN 2010da dips to $m_i\approx 23$ mag in the optical but returns to $m_i\approx 20$ mag after 500 days. In the same time frame, the IR flux declines by about one magnitude to $m_{3.6}\approx 18.7$ mag at 460 days. The optical light curve then appears to settle into an aperiodic, variable state that oscillates between $m_{r,i}\approx20$ mag and $m_i\approx22$ mag every 500$-$1000 days. The IR light curve remains roughly at its pre-eruption brightness, but the color becomes much bluer (from $m_{3.6}-m_{4.5}\approx 0.2$ to $\approx-0.2$). Beginning around $\approx$ 1500 days after discovery, the IR light curve begins to rise to magnitudes comparable to the initial outburst. We refer to the surviving star as the \textit{progeny} of SN 2010da.

\begin{figure}[t]
\centering
\includegraphics[width=0.5\textwidth]{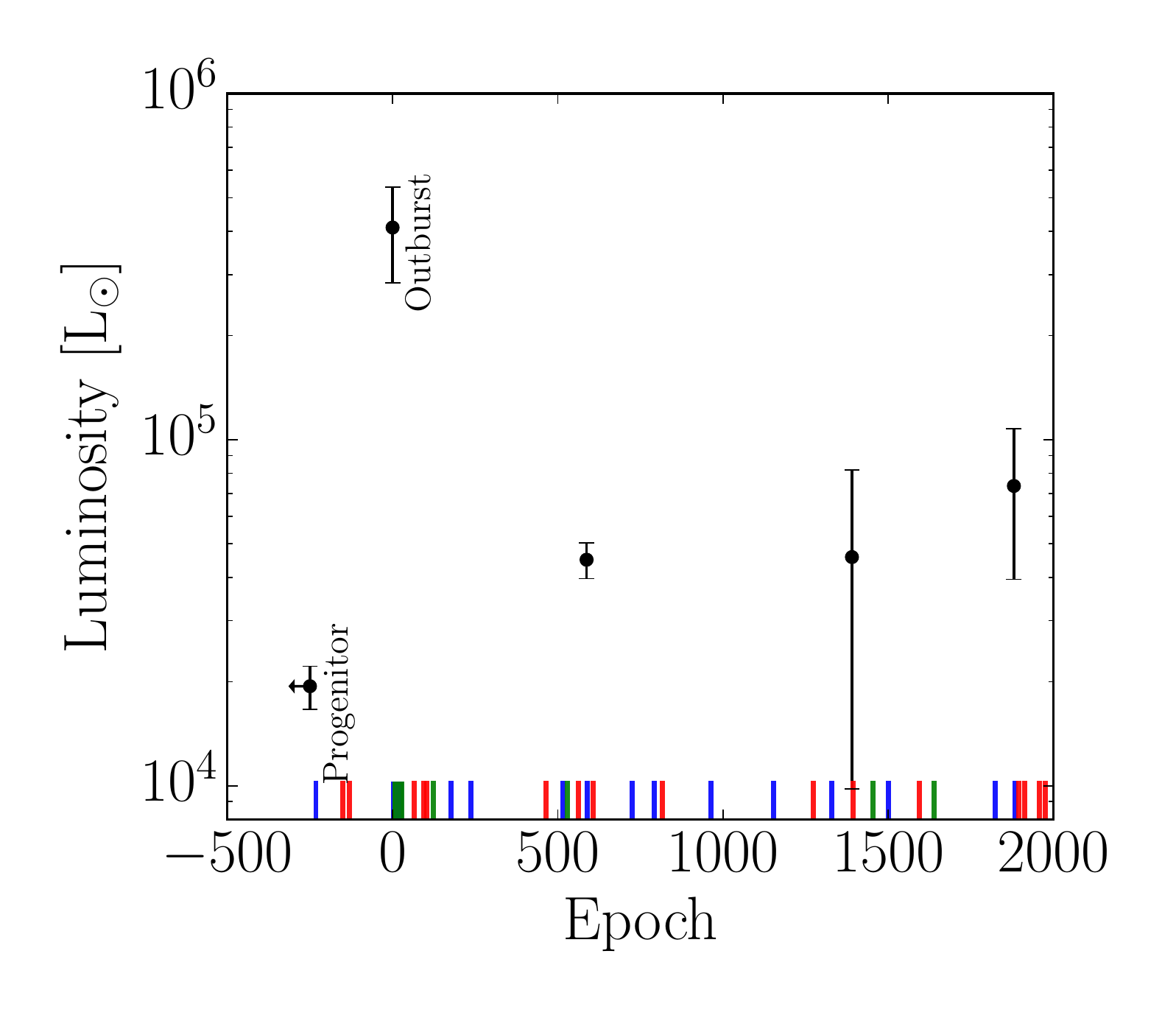}
\caption{Bolometric luminosity of SN 2010da as derived from the NIR/IR flux as a function of time (black). The colored lines at the bottom of the plot indicate our photometric X-ray (green), optical (blue), and infrared (red) coverage for reference. The progeny of SN 2010da system has a consistently higher bolometric luminosity than its progenitor.}
\label{fig:bol_lum}
\end{figure}

The progeny's optical/IR SED can be roughly described by a single blackbody with variable excess flux in the optical. After 500 days, the NIR and IR fluxes are fit by a blackbody with a radius of $\approx$ 6 AU and a temperature of $\approx 2000$ K. The derived radius is smaller than the progenitor radius at $\approx 10$ AU, and the temperature is higher than the blackbody temperature of the pre-eruption SED ($T=1500$ K), consistent with the color shift seen in the IR. The optical flux, however, varies by $\approx 2$ mag even two years after the initial outburst. Fitting our NIR/IR measurements to blackbodies, we track the bolometric luminosity of the system over time, as shown in Figure \ref{fig:bol_lum}. The luminosity of the progeny and its surrounding environment is about $2-5$ times larger than the progenitor of SN 2010da excluding contribution from the UV/optical, which supplies $\sim10-20$\% of the total luminosity.

\begin{figure*}[ht]
\centering
\includegraphics[width=\textwidth]{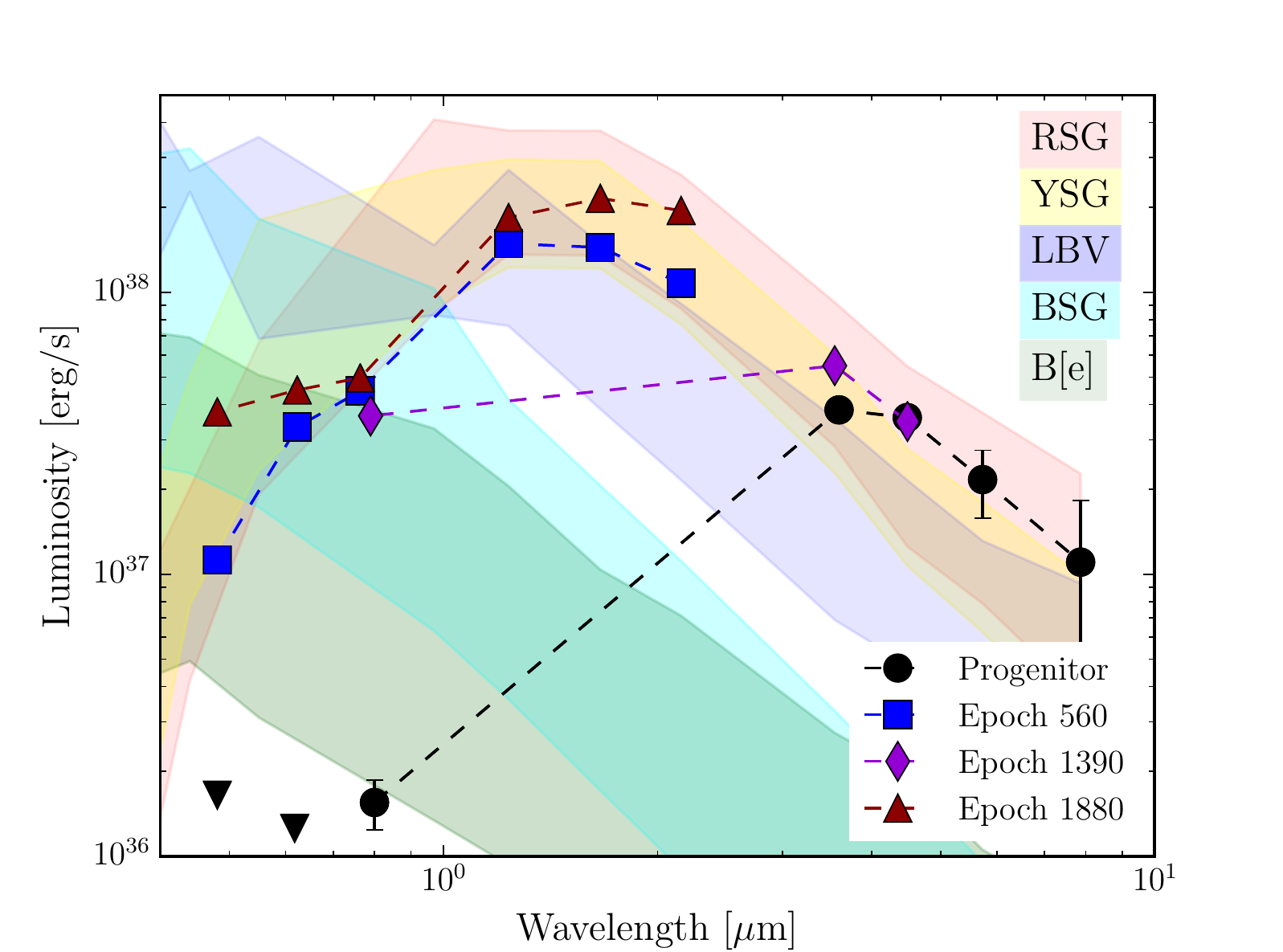}
\caption{Spectral energy distributions of SN 2010da compared to SEDs calculated using massive stars in the LMC \citep{bonanos2010spitzer}. The red region is the typical range for a RSG (M-type), yellow for a YSG (K-type), cyan for a BSG (B-type), dark blue for a LBV and dark green for a B[e] star. Although SN 2010da displays B[e] properties in its spectrum, its SED seems to follow that of a RSG or YSG. Downward facing triangles are 3$\sigma$ upper limits.}
\label{fig:sed_boronos}
\end{figure*}

We compare the SEDs taken more than 800 days after the initial outburst to a variety of SEDs of massive stars in the LMC analyzed by \cite{bonanos2009spitzer}. We group these massive stars by their spectroscopic classification reported by \cite{bonanos2009spitzer}, and we construct ``typical" SED ranges for each class using the 10$^{\mathrm{th}}$ and 90$^{\mathrm{th}}$ percentile magnitude of each filter within each group. The SEDs for red, yellow and blue supergiants (RSGs, YSGs, BSGs), luminous blue variables (LBVs) and B[e] stars compared to the SED of the progeny of SN 2010da are shown in Figure \ref{fig:sed_boronos}. Here, we are defining B[e] stars as any star with B[e] emission lines (e.g. Hydrogen Balmer, iron, etc.), regardless of luminosity class. The progeny's SED most closely matches the SED of a typical RSG. As a test, we also convert our SDSS bandpasses into Johnson magnitudes and search for the nearest neighbor of the progeny SED within the space of the magnitudes used by \cite{bonanos2009spitzer}; the nearest neighbor is [SP77]46-31-999, an M2 Iab star. The fact that the SED of the progeny most closely resembles that of a RSG does not necessarily imply that the progenitor or progeny is a RSG. In fact, the small radius we infer from the \textit{Swift}/UVOT data ($\lesssim0.5 AU$) rules out most RSG candidates. Both broadband photometry and spectroscopy are necessary when classifying obscured, massive stars.

\subsection{Spectroscopic Properties of SN 2010da}
\begin{figure*}[ht!]
\begin{center}
\includegraphics[width=\textwidth]{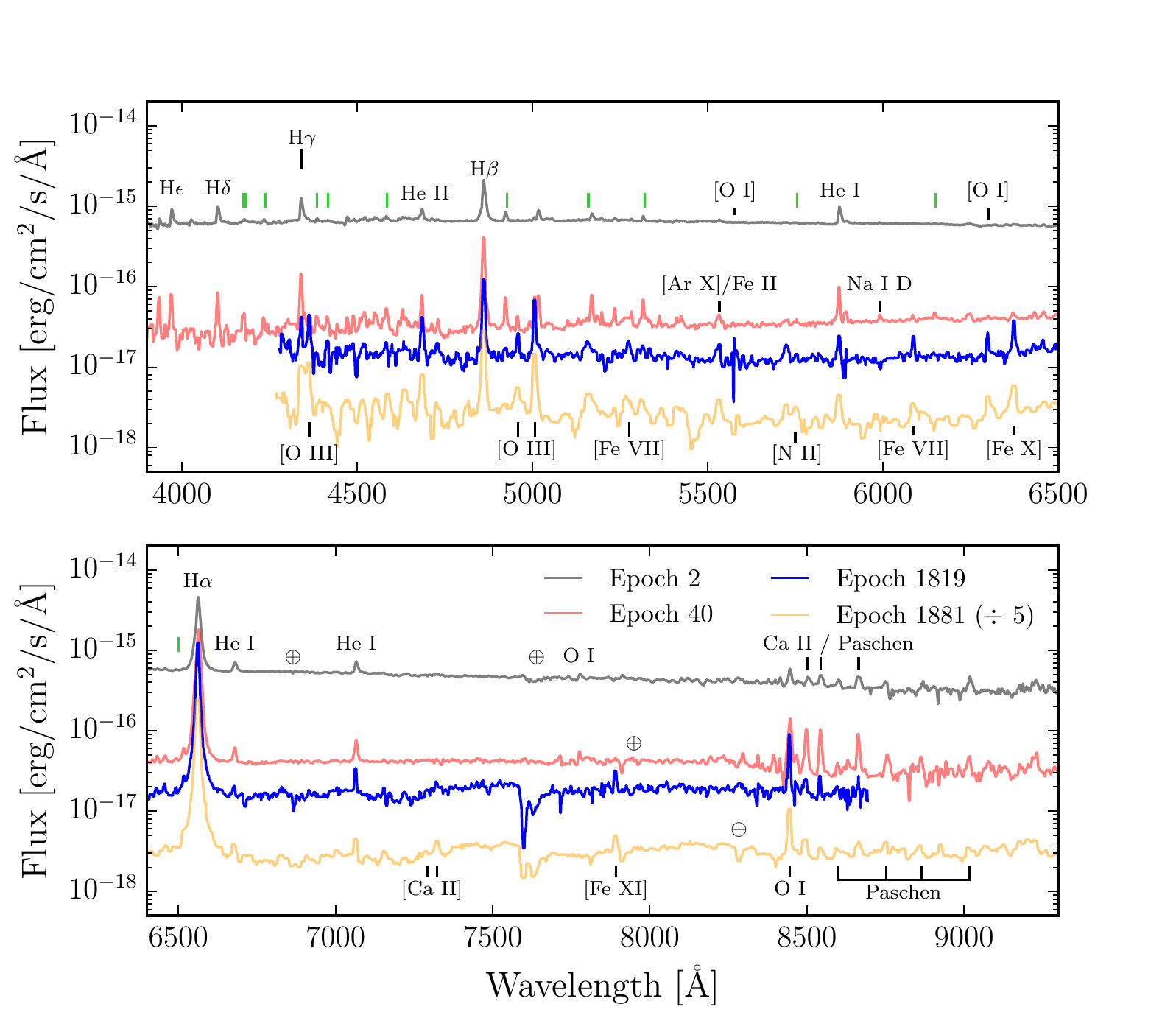}
\caption{Low-/medium-resolution optical spectra of SN 2010da. All spectra have been smoothed to a $\approx$ 10\AA\ resolution, and the 1881-day spectrum has been shifted downward by a factor of 5 for clarity. Unlabeled green lines refer to Fe II emission lines with a 3$\sigma$ detection in at least one epoch. } 
\label{fig:spectral_evolution}
\end{center}
\end{figure*}

\begin{figure*}[h!]
\begin{center}
\includegraphics[width=\textwidth]{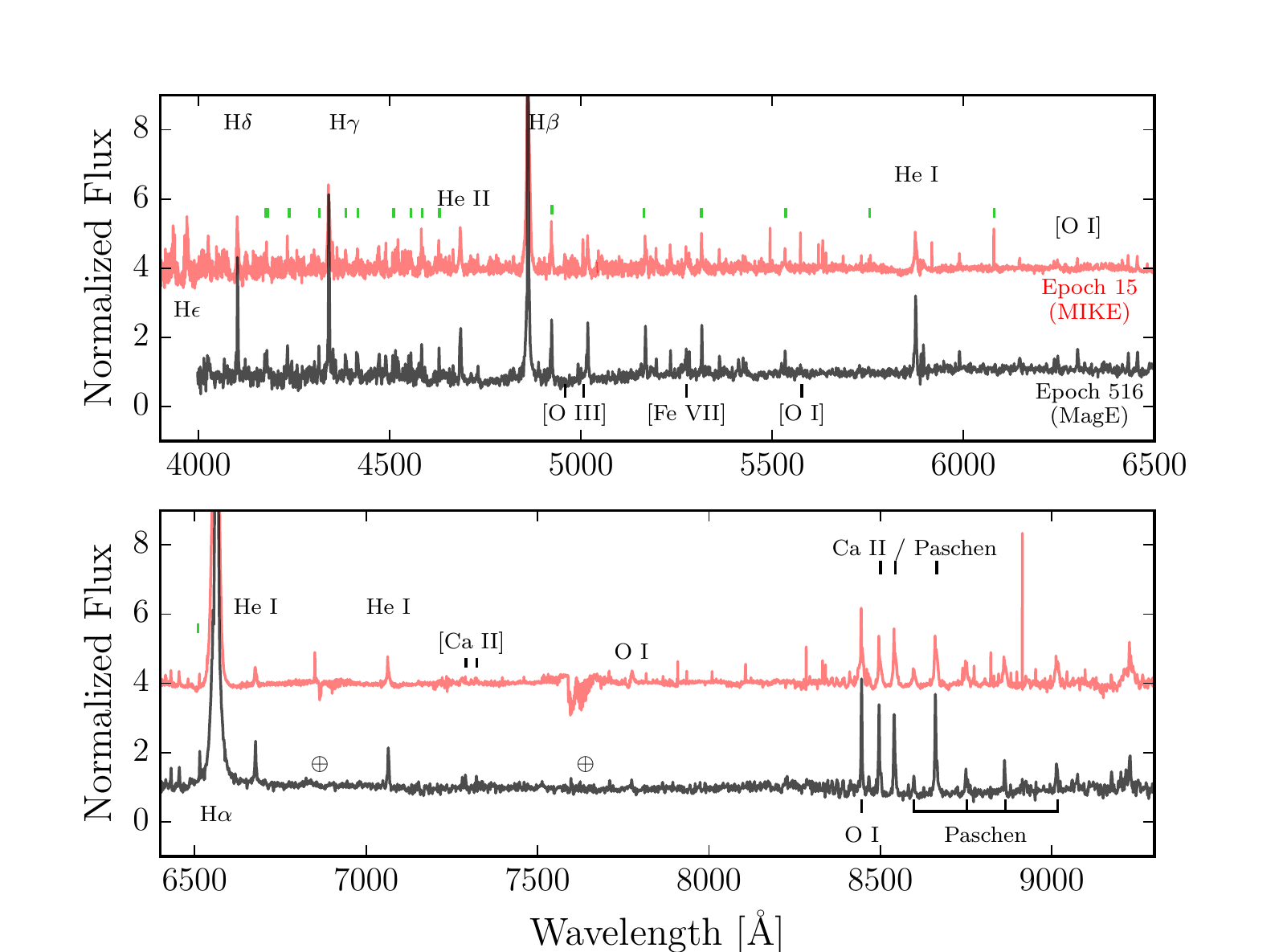}
\caption{High-resolution optical spectra of SN 2010da from MIKE and MagE. Both spectra have been smoothed to $\approx 4$\AA\ resolution for display purposes. Unlabeled green lines refer to Fe II emission lines with a 3$\sigma$ detection in at least one epoch.} 
\label{fig:high_res_spec}
\end{center}
\end{figure*}

Throughout our observations, spanning from 2 to 1881 days after the optical discovery, the spectra of SN 2010da exhibit strong hydrogen Balmer and Paschen, He I and II, Fe II and Ca II emission lines. Early spectra reveal P-Cygni profiles in the Balmer, Paschen and helium lines, while later spectra  develop strong nebular emission lines. A full list of these lines with a 3$\sigma$ detection in at least one epoch and their properties is provided in Table \ref{tab:specdetails}. The low-resolution spectra are shown in Figure  \ref{fig:spectral_evolution}, and the high-resolution spectra are shown in Figure \ref{fig:high_res_spec}. The high-resolution spectra have been normalized by fitting a low-order polynomial to the smoothed spectra. The strong Balmer lines, low excitation emission lines (especially Fe II), the forbidden lines and the IR excess all indicate that the progeny of SN 2010da exhibits B[e] phenomena by the criteria enumerated in \citealt{lamers1998improved}. This classification scheme is purely observational but is linked to a complex CSM surrounding the star (see \citealt{lamers1998improved} Section 2.2). We observe the development of high ionization emission lines of iron at later epochs and continuous He II 4686\AA\ emission. Both of these observations are due to the presence of a hard radiation field (UV/X-ray emission) associated with the HMXB nature of the object (see Section \ref{sec:hmxb}).

\subsubsection{Hydrogen Balmer Lines}
\begin{figure}[h]
\begin{center}
\includegraphics[width=0.5\textwidth]{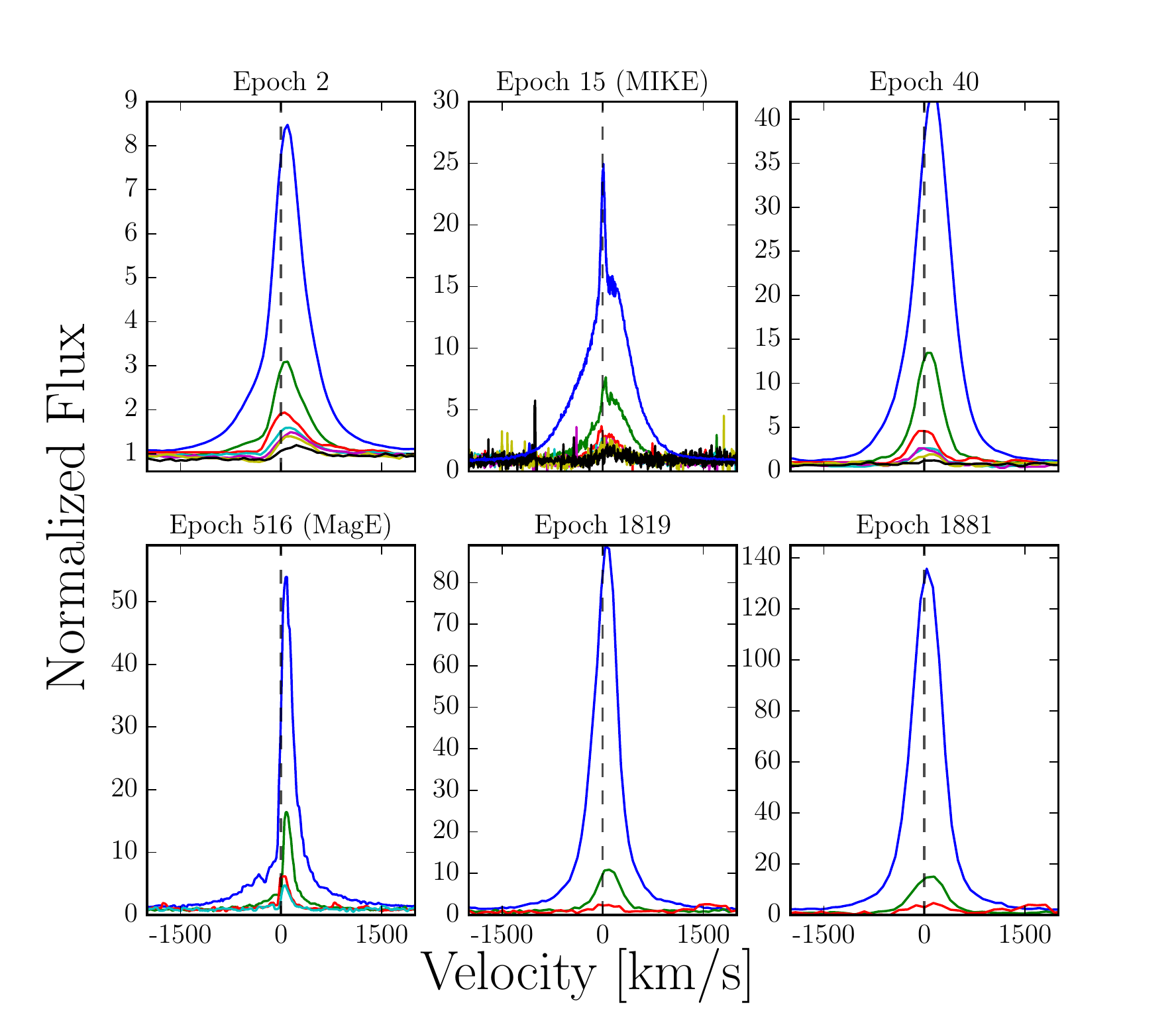}
\caption{Evolution of Balmer lines in the optical spectra. Each epoch is listed above its corresponding subplot. H$\alpha$, H$\beta$, H$\gamma$, H$\delta$, H$\epsilon$,H$\zeta$ and H$\eta$ are shown in blue, green, red, teal, pink, yellow and black, respectively. Note that the y-axis are all on independent scales for clarity. The local continuum has been normalized to one by fitting a first order polynomial.} 
\label{fig:balmer}
\end{center}
\end{figure}

The Balmer lines exhibit some of the most drastic changes of the spectrum over the span of our observations. Their equivalent widths roughly follow the optical/IR flux variations and appear to be significantly increasing in the most recent observations. A time sequence of the Balmer lines is shown in Figure \ref{fig:balmer}.

The spectra are marked by large H$\alpha$ luminosity, contributing up to $\approx$ 30\% of the $r$-band flux at later times. Following the initial outburst, H$\alpha$ is well described by a Lorentzian profile with a full width at half maximum of $\approx560$ km s$^{-1}$. The full-width at continuum intensity is $\approx 3600$ km s$^{-1}$. The H$\alpha$ flux immediately following the discovery is $\approx 6.6 \times 10^{-13}$ erg s${^{-1}}$ cm$^{-2}$\AA$^{-1}$ and approximately halves 40 days later. As the object cools, the continuum flux decreases while the H$\alpha$ flux remains relatively constant at $\approx (2-3) \times 10^{-13}$ erg s$^{-1}$ cm$^{-2}$ . At the same time, $F_{\mathrm{H}_\alpha}/F_{\mathrm{H}_\beta}$ increases from $\approx 4$ to $\approx 8$ in the first 40 days. This is far greater than the expected value of $F_{\mathrm{H}_\alpha}/F_{\mathrm{H}_\beta}\approx 2.8$ for Case B recombination at $\sim10^4$ K, the approximate temperature of the hotter blackbody component in the SED during outburst. While dust extinction may account for this excess, the continuum is unabsorbed. An alternative possibility is that a high-density CSM affects the Balmer decrement via a self absorption and collisions \citep{drake1980}. Using the static slab model at $10^4$ K from \cite{drake1980}, we find that the observed $F_{\mathrm{H}_\alpha}/F_{\mathrm{H}_\beta}$ ratio roughly corresponds to a density of $n_e \sim 10^{10}-10^{12}$ cm$^{-3}$.  At these densities, $F_{\mathrm{H}_\gamma}/F_{\mathrm{H}_\beta}$ is suppressed to $\approx 0.3$, $F_{\mathrm{H}_\delta}/F_{\mathrm{H}_\beta}$ to $\approx 0.2$, and $F_{\mathrm{H}_\epsilon}/F_{\mathrm{H}_\beta}$ to $\approx 0.15$. The observed line fluxes roughly match these predictions during the initial outburst. At 40 days later, the $F_{\mathrm{H}_\alpha}/F_{\mathrm{H}_\beta}$ ratio remains consistent with $n_e \sim 10^{10}-10^{12}$ cm$^{-3}$.

\begin{figure}[h]
\centering
\includegraphics[width=0.5\textwidth]{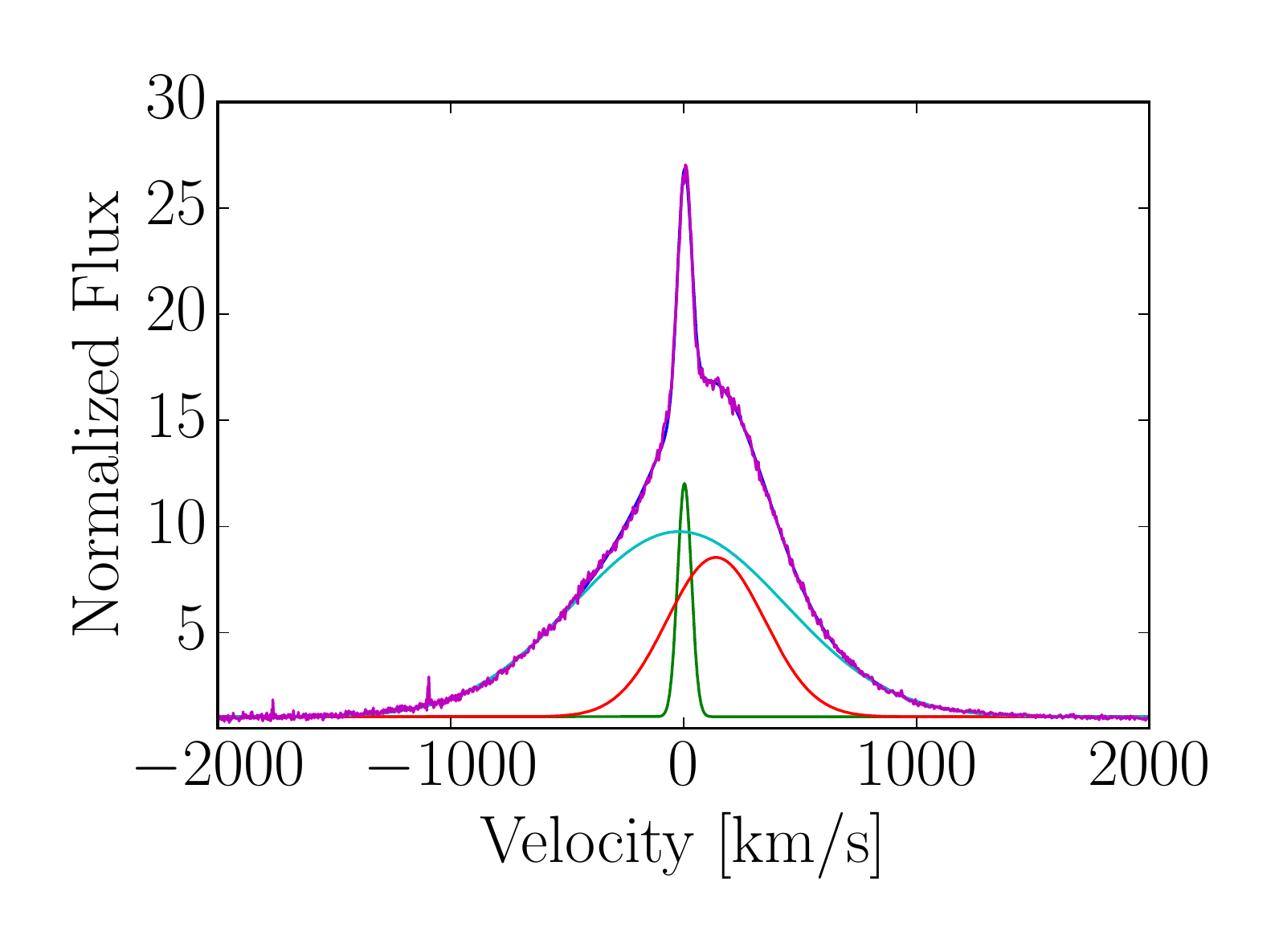}
\caption{The H$\alpha$ emission line 14 days after the outburst. The line is well described by three Gaussian components (shown in cyan, red and green). The local continuum has been normalized to one by fitting a first order polynomial.}
\label{fig:Halphafit}
\end{figure}

In our high-resolution MIKE spectrum taken 14 days after discovery, the H$\alpha$ line includes multiple components. We fit the H$\alpha$ profile with three Gaussian components: a narrow component (FWHM $\approx70$ km s$^{-1}$), an intermediate component (FWHM $\approx500$ km s$^{-1}$) and a broad component (FWHM $\approx 1060$ km s$^{-1}$)  with $\chi^2_r\approx1.9$ (see Figure \ref{fig:Halphafit}). Multi-component (specifically three-component) lines are common in dusty ILOTs \citep{berger2009,van2012s,tartaglia2015,turatto1993}. The narrow component is broader than the other detected narrow emission lines (e.g. other Balmer lines with FWHM $\sim40-60$ km s$^{-1}$) possibly due to electron scattering. These narrow components are consistent with a pre-existing wind, possibly from an earlier red supergiant phase. As with SNe IIn emission lines, the intermediate component is ascribed to the shockwave-CSM interaction (see \citealt{chevalier1994}), although the velocity is nearly an order of magnitude slower than in SNe. The intermediate component is significantly red-shifted (by $\approx 140$ km s$^{-1}$) relative to the narrow component. The apparent redshift may be an artifact of electron-scattering through high optical depths and is often seen in other dusty ILOTs, giants, Wolf-Rayet stars and other stars experiencing significant mass loss \citep{humphreys2011photometric,hillier1991}. The broadest component is only identified in H$\alpha$, which may be due to lower signal-to-noise in the other lines or additional scattering. The central wavelength of this component is consistent with that of the intermediate component, also suggesting a common physical origin.

At later times, the H$\alpha$ emission line can be described as roughly Gaussian or Lorentzian with an extended red wing and a FWHM $\sim 300-600$ km s$^{-1}$. This red wing is especially apparent in the high-resolution MagE spectrum at 516 days. The FWHM of the late-time H$\alpha$ emission is consistent with the intermediate component of the H$\alpha$ line during the initial eruption, while the narrow component remains unresolved in all other spectra. This suggests that the late-time emission is powered by a persistent wind or mass loss consistent with that of a blue or yellow supergiant.

\subsubsection{Ca II Lines}
\begin{figure}[h]
\centering
\includegraphics[width=0.45\textwidth]{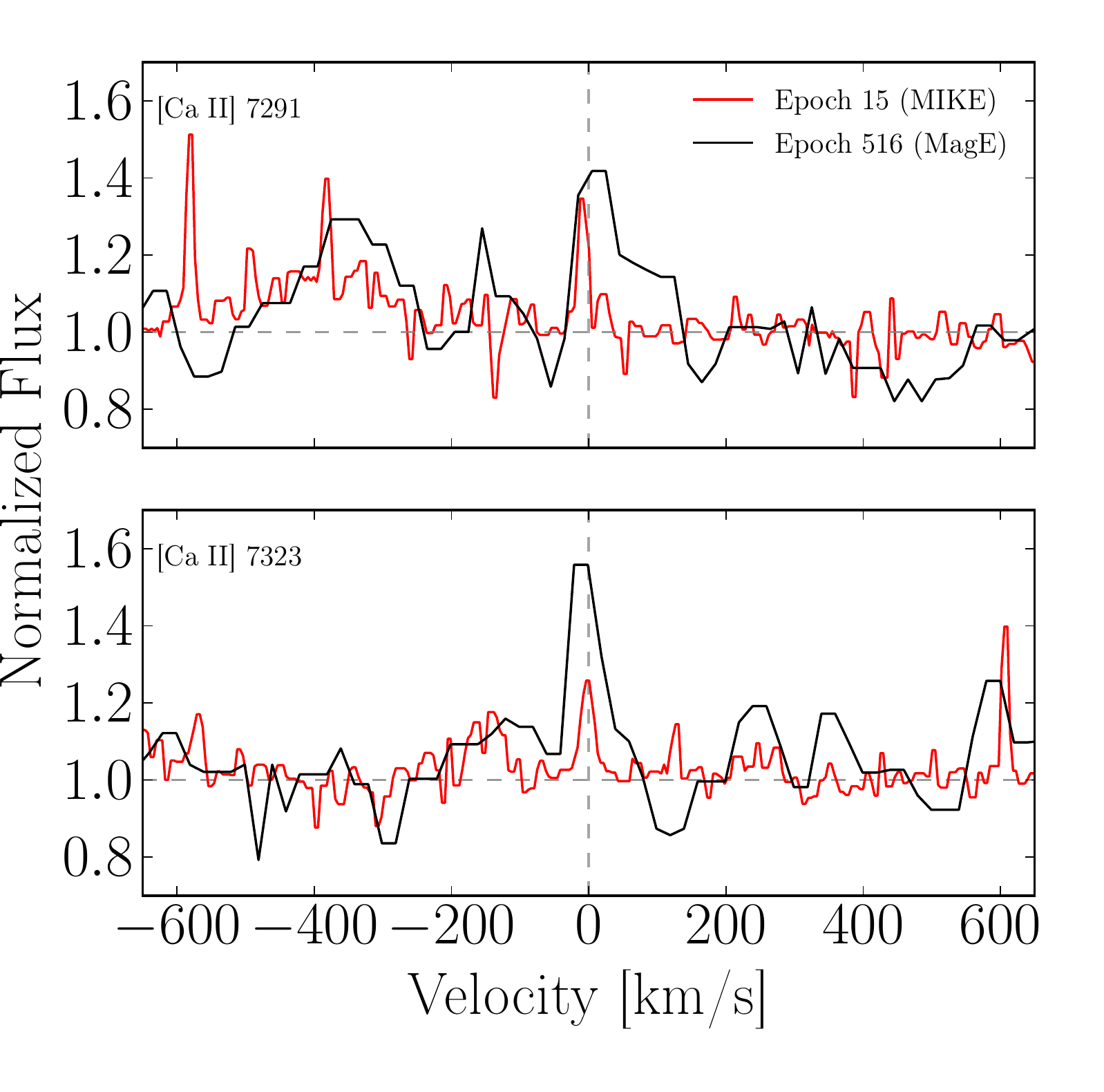}
\caption{Narrow, [Ca II] lines detected in our high-resolution spectra. The local continuum has been normalized to one by fitting a first order polynomial.}
\label{fig:caii}
\end{figure}

Narrow [Ca II] lines (FWHM $\lesssim$50 km s$^{-1}$) are detected in our highest resolution (MIKE) spectrum 15 days after discovery and possibly again in the MagE spectrum (Figure \ref{fig:caii}). Similar forbidden calcium emission was observed in NGC\,300 OT2008-1 \citep{berger2009} and SN 2008S \citep{botticella2009sn}, as well as in several warm hypergiants \citep{humphreys2013}, and its presence is typically associated with a dusty environment. Because collisional de-excitation normally drives calcium to its ground state, the [Ca II] doublet is associated with cooler, low density regions. Forbidden calcium lines are additionally suppressed by UV radiation due to the low ionization potential of calcium. Highly ionized iron and He II lines indicate a strong UV radiation field. We can conclude from this fact and the narrow line shape that the [Ca II] forbidden lines are from excited calcium located in the outer CSM, likely in the original dust shell at $\approx 10$ AU.

The presence of the calcium triplet also supports the existence of a cool, low density circumstellar environment \citep{polidan1976spectroscopic}. However, the calcium triplet is significantly blended with hydrogen Paschen emission, so we cannot make a definitive statement about the line shape or strength. The Ca H $\&$ K  doublet, typically associated with the calcium triplet, are also blended with H$\eta$ and an iron line.

\subsubsection{Fe Lines}
We detect strong Fe II emission lines in all spectra. The Fe II features roughly match the hydrogen Balmer series in shape, FWHM and line offset, indicating that these features also arise from material within the ejecta and CSM. Fe II emission is seen in NGC300 OT2008-1 and M85 OT2006-1 \citep{berger2009,bond20092008}, although the lines seen in SN 2010da are notably stronger.

\begin{figure}[t!]
\begin{center}
\includegraphics[width=0.45\textwidth]{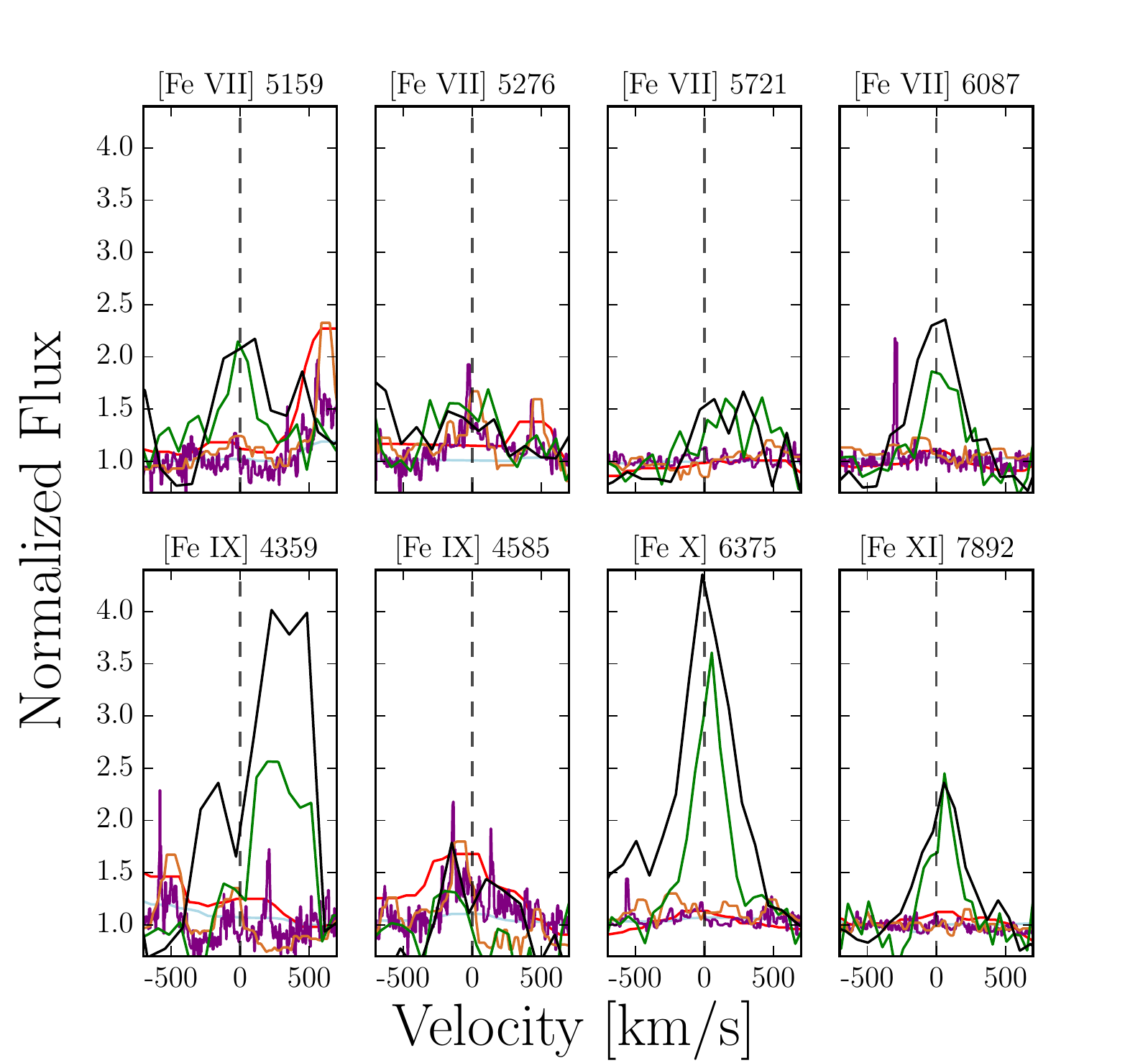}
\caption{Evolution of the coronal iron lines. Spectra taken on days 2, 15, 40, 516, 1819 and 1881 are shown in light blue, purple, red, orange, green, and black, respectively. The local continuum has been normalized to one by fitting a first order polynomial.} 
\label{fig:iron_spectral}
\end{center}
\end{figure}

In addition to Fe II, we detect emission from high ionization, forbidden iron lines, including [Fe VII], [Fe X] and [Fe XI] in the last two epochs of spectroscopy (Epochs 1819 and 1881; see Figure \ref{fig:iron_spectral}). These forbidden iron lines are not typically seen in ILOTs due to their weakly ionizing radiation. High ionization iron lines are occasionally found in SNe IIn such as SN 1997eg \citep{hoffman2008dual} and SN 2005ip \citep{smith2009coronal} due to shock heating of the surrounding CSM. Unlike SNe IIn, the iron lines seen in SN 2010da do \textit{not} arise from continual shock heating over hundreds of days. Instead, these lines arise in regions of diffuse gas surrounding the progeny which are heated to temperatures of about $2\times10^6$ K, the approximate ionization temperature for lines such as [Fe XII] and [Fe XIV] \citep{corliss1982energy}, by X-rays from the compact companion (see Section 4.3). 

\subsubsection{He Lines}
\begin{figure}[h]
\centering
\includegraphics[width=0.45\textwidth]{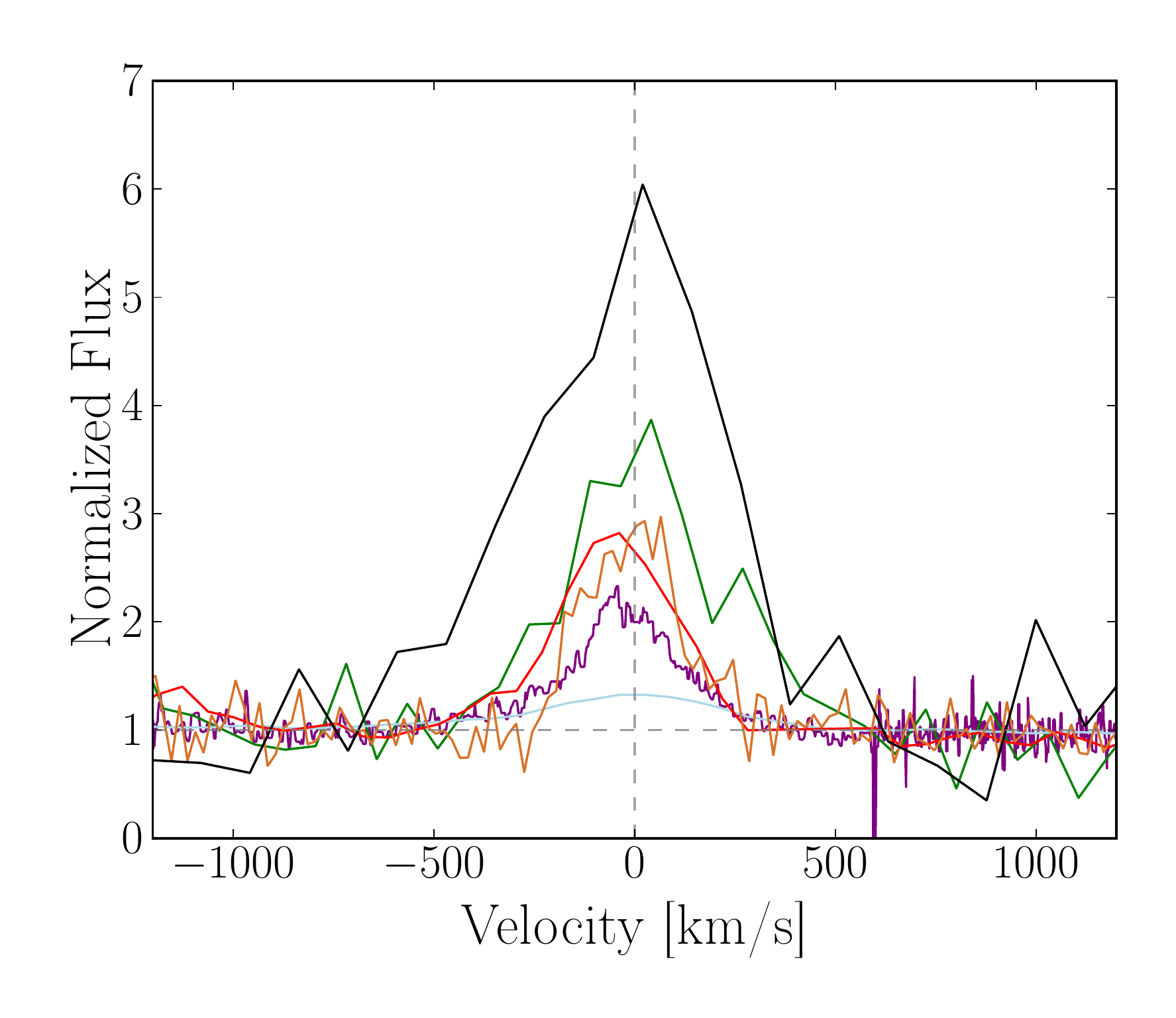}
\caption{Evolution of the He II 4686\AA\ emission. Spectra taken on days 2, 15, 40, 516, 1819 and 1881 are shown in light blue, purple, red, orange, green and black, respectively. The local continuum has been normalized to one by fitting a first order polynomial.} 
\label{fig:HeII}
\end{figure}

He I and II emission lines are seen throughout our observations, as shown in Figure \ref{fig:HeII}. The widths (FWHM $\sim200-400$ km s$^{-1}$) and profiles of the He I lines largely follow the Balmer series with a double-peaked structure in our high-resolution spectra. We additionally detect single-peaked He II 4686\AA\ emission during each epoch. The low-resolution spectra are unable to resolve the He II 4686\AA\ line, but our MIKE spectrum reveals a FWHM $\approx 270$ km s$^{-1}$. He II 4686\AA\ has a relatively high ionization potential and is sensitive to the EUV flux of the system. For this reason, it is often linked to the accretion onto a compact object \citep{lewin1997x}. The continual presence of He II 4686\AA\ emission in each of our observations is due to the compact companion and indicates consistent mass transfer onto this compact object. 

\subsubsection{Additional Absorption and Emission Features}

Hydrogen Paschen emission lines extending to approximately Pa30 are observed within the first 15 days of the outburst. The high-resolution MIKE spectrum reveals double-peaked emission with narrow and broad components, similar to the Balmer lines. These profiles are not resolved in our lower-resolution spectra. Within the first 40 days, the ratio $F(\mathrm{P}_\delta)/(\mathrm{H}_\beta)$ decreases from $\approx 0.2$ to $\approx 0.08$. The latter value is roughly consistent with that expected from Case B recombination ($\approx 0.07$), although the effect of high electron density on this line is unclear. 

At early times, we detect O I at 7774\AA\ and 8446\AA\  in emission. The O I 8446\AA\ line shows the same double-peaked profile as the Balmer series, while O I 7774\AA\ maintains a P-Cygni profile until 40 days after the initial eruption. After 1800 days, O I 7774\AA\ becomes undetectable while O I 8446\AA\ strengthens. The expected ratio between these lines is $F(\text{O I 8446\AA})/F(\text{O I 7774\AA})=0.6$, implying that O I 7774\AA\ should be detectable. The independent strengthening of O I 8446\AA\ can be attributed to Ly$\beta$ emission which is outside of our observed spectral range. Ly$\beta$ photons pump O I from the ground state to an unstable state whose decay produces O I 8446\AA\ emission \citep{mathew2012study}. This is consistent with the increased Balmer emission and UV flux at later times. In addition to O I, we detect [O I] and [O III] features. Unfortunately, the [O III] 4363\AA\ feature appears blended with either [Fe IX] or an Fe II emission line, making it difficult to use the [O III] ratios to calculate the electron temperature. 

Unresolved Na I D lines are observed in the two latest epochs (517 and 1817 days) as emission and absorption respectively. The variability of these lines indicates that they are associated with the CSM rather than interstellar medium.

\subsection{X-ray Spectral Modeling}
\label{sec:xrayspec}
We model the X-ray emission from SN 2010da and its progeny using XSPEC version 12.8.2n \citep{arnaud1996xspec}. We use the Cash statistic, a derivative of the Poisson likelihood, as our fit statistic. To test our fits, we use the XSPEC built-in command {\tt goodness} to perform $10^4$ Monte Carlo simulations of the spectral data. For each simulation, the program calculates the Cram\'er von Mises (CVM) test statistic, which is shown to be a good fit statistic for the data derived from a Poisson distribution \citep{spinelli1997cramer}. If about 50\% of these simulations have a CVM statistic less than that of our model, the best-fit model is a good representation of the data. A percentage much lower than 50\% implies that our model is over parametrized, and a percentage much greater than 50\% implies that our model is inconsistent with the data. All reported errors correspond to $1\sigma$ error bars (the 68\% confidence interval).

We combine all of the \textit{Swift}/XRT $0.5-8$ keV data taken within 40 days of the outburst and fit it to a power law with Galactic absorption ({\tt tbabs * pow}) with $N_\mathrm{H,MW}= 4\times10^{20}$ cm$^{-2}$. We find that additional absorption over-parametrizes the model ({\tt goodness} $=15$\%), but an excess column density as large as $N_H\approx5\times10^{21}$ cm$^{-2}$ is consistent with the data. Our best fit model is described by $\Gamma = -0.05^{+0.11}_{-0.10}$ with an unabsorbed $0.3-10.0$ keV flux of $9.62^{+0.87}_{-0.85} \times10^{-16}$ erg s$^{-1}$ cm$^{-2}$ (assuming $N_\mathrm{H} = 0$). This corresponds to a luminosity of $3.98^{+0.36}_{-0.35}\times10^{38}$ erg s$^{-1}$. Similarly, we fit the first \textit{Chandra} observation (Epoch 123) to an absorbed power law. We find the best fit model is described by $\Gamma = 0.26^{+0.20}_{-0.21}$ with an unabsorbed $0.3-8.0$ keV luminosity of $1.95^{+0.17}_{-0.48}\times10^{37}$ erg s$^{-1}$. We find an absorption upper limit beyond the Galactic column of $N_H \lesssim 4 \times10^{21}$ cm$^{-2}$. The estimated column density and the photon index are degenerate such that a higher column density implies a softer power law.

\begin{figure}[t!]
\centering
\includegraphics[width=0.45\textwidth]{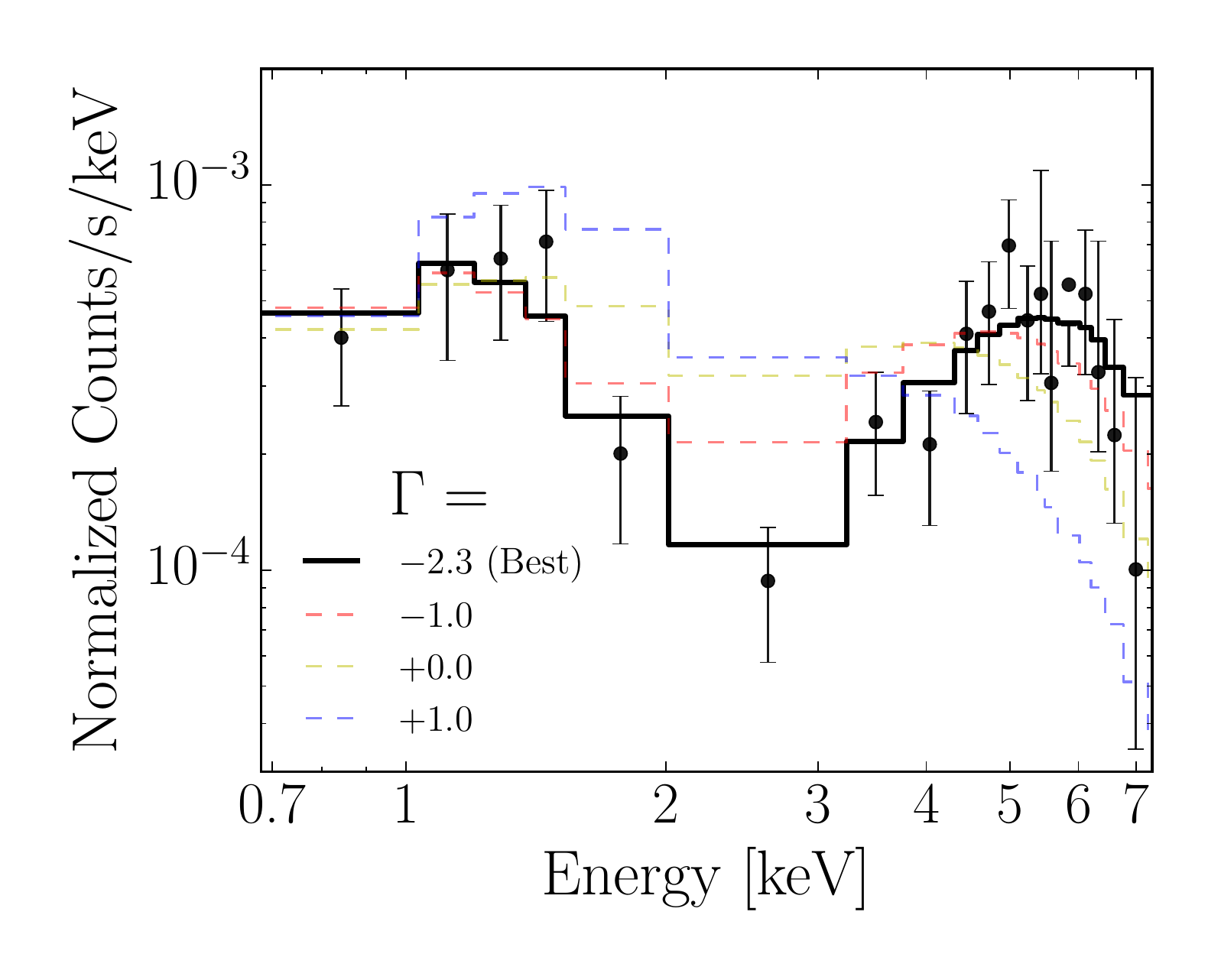}
\caption{The latest X-ray spectrum of SN 2010da from \textit{Chandra} Observation 16029, normalized by the detector's effective area. Shown are four models with a Bremsstrahlung and power law component. We fix the photon index, $\Gamma$, to -1.0 (red), 0.0 (yellow) and +1.0 (blue) and compare to the best fit model with $\Gamma = -2.3$. We are unable to recover the bimodal structure of the spectrum with softer power laws. }
\label{fig:pow_laws}
\end{figure}

\begin{figure}[ht!]
\centering
\includegraphics[width=0.45\textwidth]{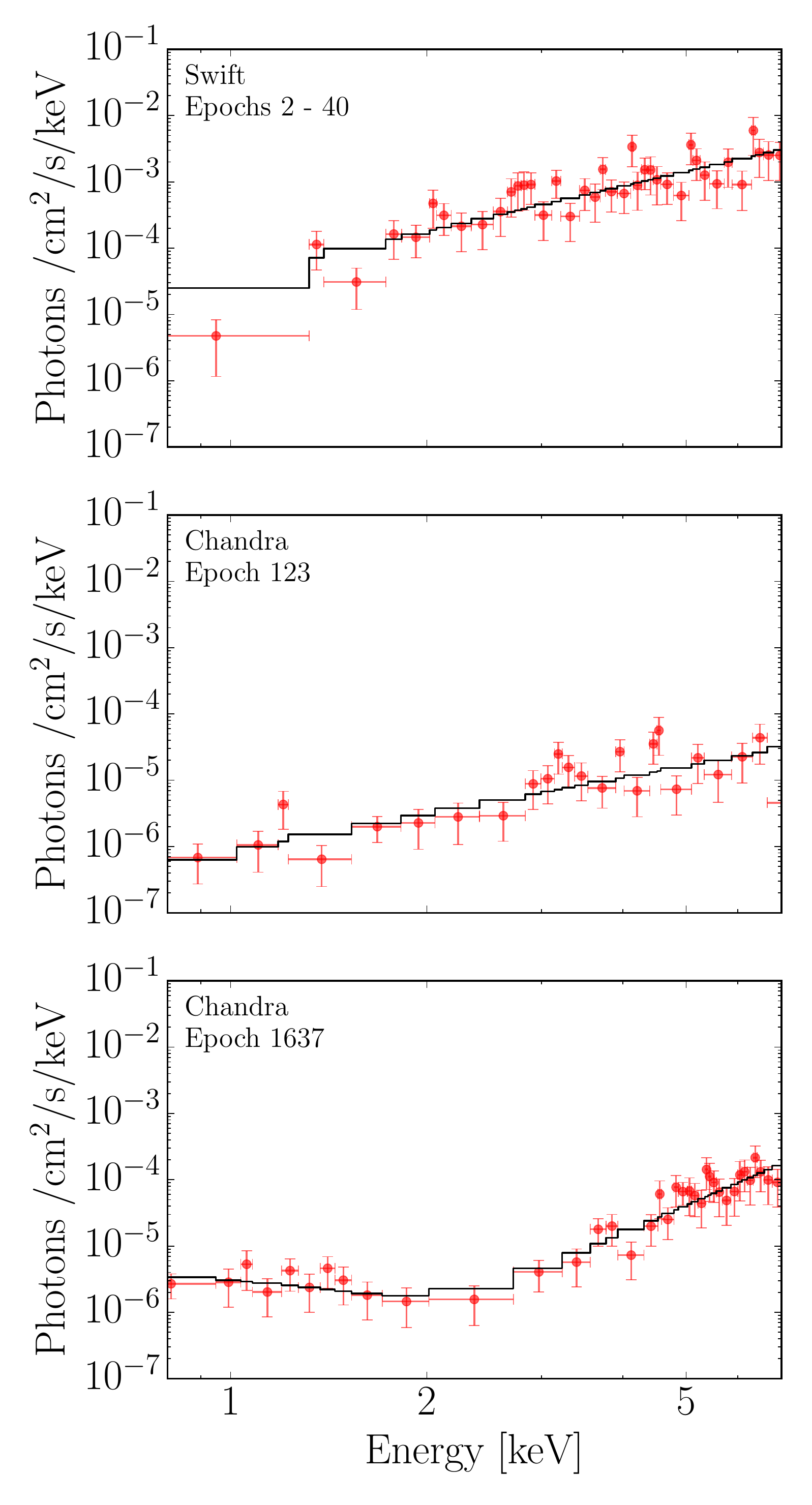}
\caption{Top: The \textit{Swift} X-ray spectrum created using data taken within 40 days of the transient. The best fit power law model is overlaid. Middle/Bottom: The \textit{Chandra} X-ray spectra (Obs. ID 12238 and 16029) with best fit power law and disc models (see text for details).}
\label{fig:xray_all}
\end{figure}

Due to limited statistics, we are unable to fit a spectrum to the second \textit{Chandra} observation (Epoch 1453). In the third \textit{Chandra} observation (Epoch 1638), there is a significant decrease in counts between $\approx 2-3$ keV. We are unable to fit this spectrum to a single power law or blackbody component and instead combine a power law with either a soft Bremsstrahlung ({\tt bremss}) or blackbody disk ({\tt diskbb}) model, with no statistical preference for either model based on the CVM statistic. For both models we obtain a similar power law with index $\Gamma = -2.2_{-0.5}^{+0.3}$ for the Bremsstrahlung model and $\Gamma = -2.3_{-0.4}^{+0.4}$ for the disk model. The Bremmstralung component has a temperature of  $0.6_{-0.2}^{+0.3}$ keV, while the disk model has an inner-disk temperature of $0.33_{-0.06}^{+0.08}$ keV. While these fits were performed by fixing the hydrogen column density to the Galactic value, fixing $N_H$ to values as high as $4\times10^{21}$ cm$^{-2}$ also gives reasonable (although statistically less favorable) fits with softer power laws. This hard power law differs from the recent results of \cite{binder2016}, who find $\Gamma\approx 0$. Specifically, we are unable to reconcile the double peak in the spectrum with softer power laws (see Figure \ref{fig:pow_laws}).  The extremely hard power law in our models indicates that additional and detailed modelling is necessary to explain this unusual \textit{Chandra} spectrum. All X-ray spectra and models are shown in Figure \ref{fig:xray_all}.

\subsection{The X-ray and UV Light Curves}
\begin{figure}[t!]
\centering
\includegraphics[width=0.5\textwidth]{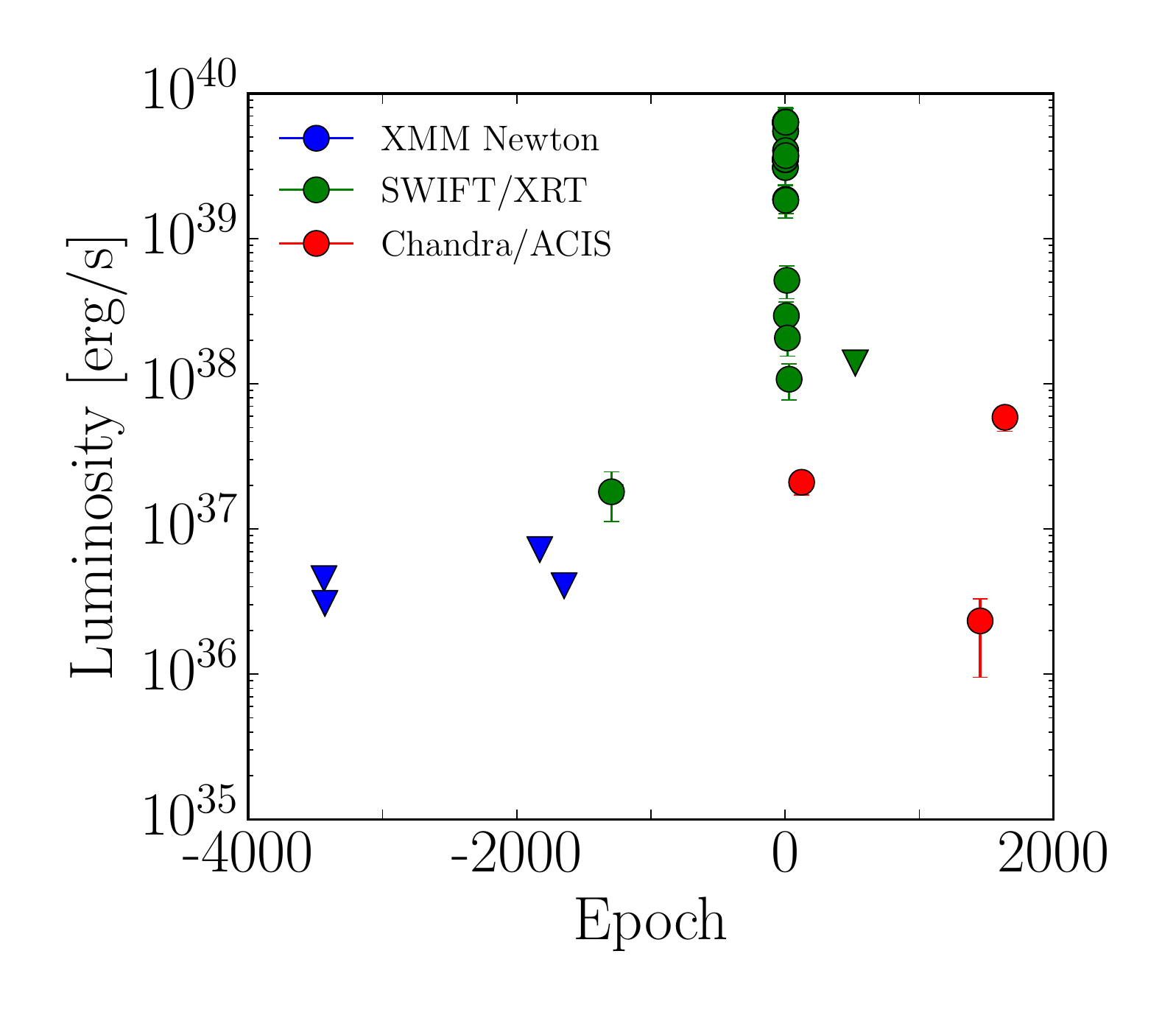}
\caption{X-ray light curve of SN 2010da. Triangles indicate $3\sigma$ upper limits. Downward facing triangles are 3$\sigma$ upper limits.}
\label{fig:xrayLC}
\end{figure}

Using archival observations of SN 2010da (Section \ref{sec:XrayObs}), we are able to construct the X-ray and UV light curves of SN 2010da. The full X-ray light curve is shown in Figure \ref{fig:xrayLC}. We build the \textit{Swift}/XRT light curve by converting the light curve produced automatically by the UK \textit{Swift} Science Data Center from a count rate to a flux using the conversion factor found for the XRT spectrum. This light curve was dynamically binned using a rate factor of 10 and a bin factor of 5.

\cite{binder2011} estimate a 3$\sigma$ upper limit on the unabsorbed $0.3-10$ keV luminosity of the progenitor to be $\approx 3\times10^{36}$ erg s$^{-1}$ using archival \textit{XMM-Newton} data. However, about 1300 days before the optical outburst, we find a weak \textit{Swift}/XRT detection at 2.6$\sigma$ with a luminosity of $1.8^{+0.7}_{-0.7}\times10^{37}$ erg s$^{-1}$, indicating X-ray variability even before the 2010 optical outburst.

\begin{figure}[t]
\centering
\includegraphics[width=0.5\textwidth]{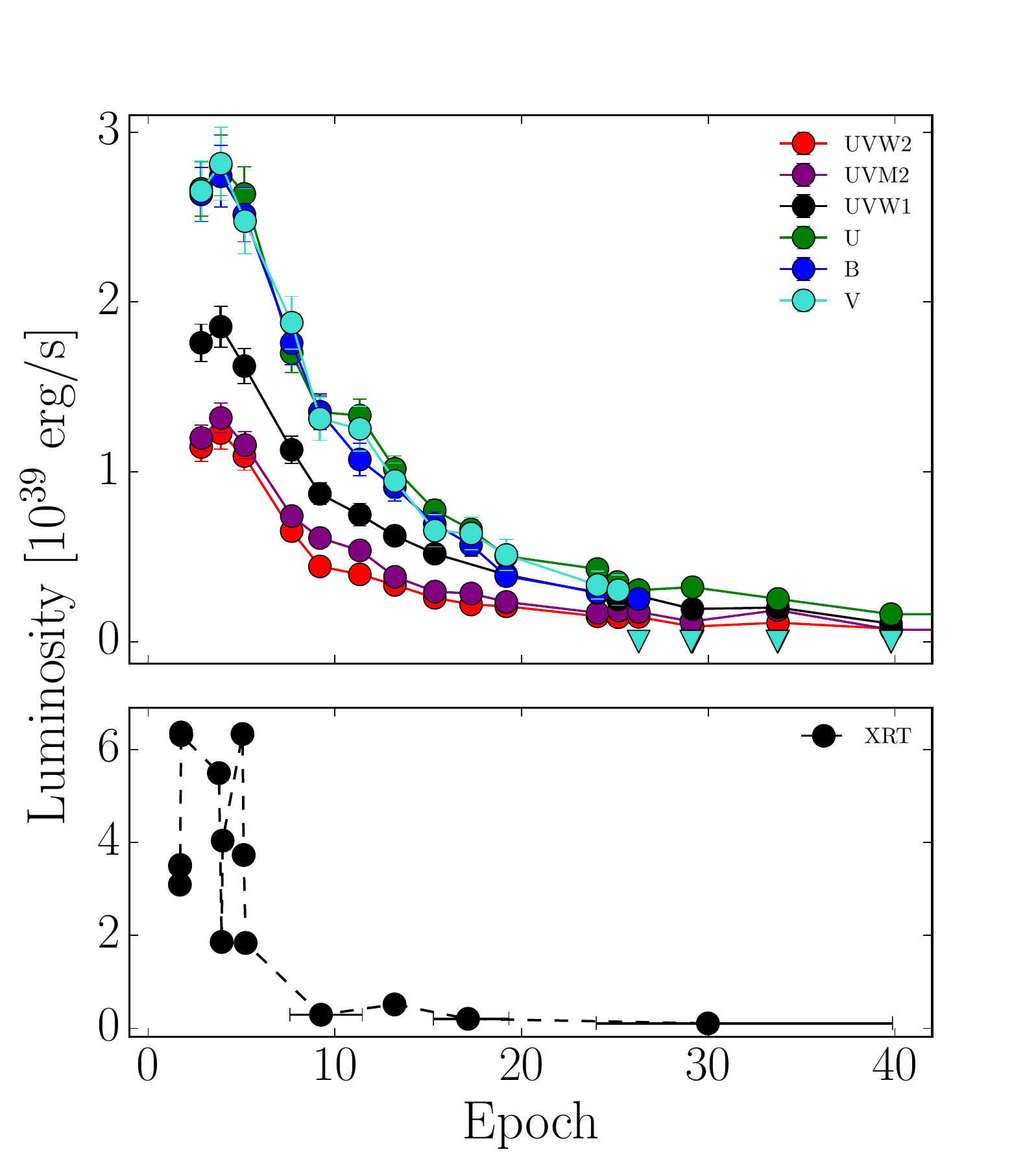}
\caption{XRT (top) and UVOT light curves near the time of discovery. Downward facing triangles are 3$\sigma$ upper limits.}
\label{fig:uvot_zoom}
\end{figure}

During the transient, the X-ray luminosity increases to a peak of $\approx 6\times10^{39}$ erg s$^{-1}$, making SN 2010da an ultraluminous X-ray source well above the Eddington limit of a 1.4 M$_\odot$ neutron star. (We note that this luminosity is larger than the luminosity reported from the spectral fit in Section \ref{sec:xrayspec} and \citealt{binder2011}, because the spectral fit averaged the luminosity over 40 days following the initial outburst.) In the week following discovery, the X-ray luminosity fluctuates between $2\times10^{39}$ erg s$^{-1}$ and $6\times10^{39}$ erg s$^{-1}$ before decaying with an e-folding time of $\approx 3.5$ days. This decay rate is slightly longer that of the UVOT light curves ($\approx 10$ days, shown in Figure \ref{fig:uvot_zoom}) and is much shorter than the decay rates found in the eruptions of $\eta$ Car ($\approx200$ days, from \citealt{binder2011}). About 1450 days after the transient, we find an X-ray luminosity of $\approx 2.4\times10^{36}$ erg s$^{-1}$ which increases to $\approx 5.9\times10^{37}$ erg s$^{-1}$ at about 1640 days. This increase in X-ray luminosity occurs at roughly the same time as the increase in optical/IR luminosity.

\section{Discussion}

Taken together, the X-ray, optical and IR light curves and spectra consistently describe an HMXB undergoing an episode of active accretion which is fueled by persistent eruptions of the primary star, with SN 2010da representing the largest observed eruption in nearly a decade of observations. The optical and IR light curves are powered by mass loss of the supergiant. This mass accretes onto the compact object, giving rise to X-ray emission. The X-rays in turn excite the high ionization He II and coronal iron lines seen in the optical spectra. In this section, we summarize the properties of the progenitor and the surviving progeny of the 2010 eruption, and we discuss potential compact companions.

\subsection{The Progenitor of SN 2010da}

Ignoring any contribution from a compact companion or accretion disk, our blackbody fit of the progenitor SED (with a temperature of 1500 K and a radius of 9.6 AU) reveals a stellar bolometric luminosity of about $2\times10^4$ L$_\odot$. This luminosity is consistent with a 15 M$_\odot$ main sequence star \citep{meynet2000} or a supergiant with a 10$-$12 M$_{\odot}$ ZAMS mass \citep{ekstrom2012}. The low temperature suggests that this blackbody is not the photosphere of the progenitor. Instead, we interpret this SED as a dusty shell surrounding the star. 

To further investigate progenitor candidates, we model the dusty environment of the progenitor and its SED using the radiative transfer code {\tt DUSTY} \citep{ivezic1999user}. {\tt DUSTY} is able to model the density profiles of spherically symmetric, radiatively driven winds, requiring as input the central source SED, the dust composition, the optical depth and the inner dust temperature. Since we do not see silicate features around 8 $\mu$m in our pre-eruption \textit{Spitzer} observations, we choose a pure graphite environment \citep{drake1980}. The carbon-rich dust is consistent with the stability of the dust shell at a relatively high temperature ($\approx 1500$ K), which has a higher sublimation temperature than silicate \citep{kobayashi2011}. We assume that the shell has a thickness of $R_{\mathrm{out}}/R_\mathrm{in} = 2$ and use a power law density model which falls off as $\rho\propto r^{-2}$, typical of a wind. We additionally assume that the central source is a blackbody, and we leave its temperature as a free parameter. The final luminosity of the model is calculated using the normalized flux and radius computed by {\tt DUSTY}. The UVOT observations during the 2010 outburst constrain the progenitor radius to be $\approx 120$ R${_\odot}$. This limits the progenitor temperature to $T\gtrsim6200$ K. We are additionally unable to find satisfactory fits ($\chi^2_r < 2$) of the progenitor SED for temperatures above $\approx 18,000$ K. The temperature of a 15 M$_\odot$ main sequence star is about 30,000 K, meaning that we can rule out such a progenitor. Due to the low luminosity, we can also rule out an LBV progenitor, which was previously suggested by others \citep{binder2016}. The only remaining viable option at this luminosity is an evolved yellow or blue supergiant progenitor. 

We can additionally use the {\tt DUSTY} models to estimate the mass loss rate of the progenitor. Following \cite{kochanek2012unmasking}, the mass loss is approximately equal to: $$ \dot{M}\approx\frac{\kappa_V}{8\pi v_w R_\mathrm{in}} $$ where the opacity is $\kappa_V\approx 120$ cm$^{2}$ g$^{-1}$, we assume a wind velocity $v_w\approx 40$ km s$^{-1}$ (the approximate line width of the narrow Balmer/He lines from the high resolution MIKE spectrum), and $R_\mathrm{in}$ is the inner radius of the dusty shell as calculated by {\tt DUSTY}. For the range of plausible models, the estimated mass loss rates are $(4-5)\times10^{-7}$ M$_\odot$ yr$^{-1}$. This is in agreement with typical mass loss rates of RSGs of this luminosity, significantly smaller than in super-AGB stars \citep{mauron2011mass,poelarends2008supernova} and greater than in BSGs \citep{martins2015mass}. However, asymmetry and inhomogeneity (e.g. clumpiness) in the CSM can greatly affect our estimated optical depth. A more extensive review of these effects can be found in \cite{kochanek2012unmasking}.

\subsection{The Progeny of SN 2010da and Its Environment}

Our extensive photometric and spectroscopic datasets indicate that the source of SN 2010da is still active and underwent a dramatic transition to a bluer and hotter SED with a smaller radius of $\approx 6$ AU after the 2010 eruption. Additionally, the progeny is significantly more luminous than the progenitor by a factor of $\sim 2-5$. Although it is possible that the bolometric luminosity of the progenitor was larger than we predict with a significant fraction of light contributed at longer wavelengths from cool dust which was heated during the transient, it is most likely that the ongoing mass ejections and their interaction with a compact companion/CSM are injecting additional energy into the system. 

In addition to being brighter, the source is also undergoing significant variability in the optical of $\approx 1-2$ mags within a few hundred days. The variability and bolometric magnitude of the progeny (M$_\mathrm{bol}\approx -7$) are reminiscent of supergiant long-period variables \citep{wood1983}, although these do not typically show B[e]-like emission lines in their spectra nor are they often surrounded by a thick CSM. 

\begin{figure*}[h]
\centering
\includegraphics[width=0.8\textwidth]{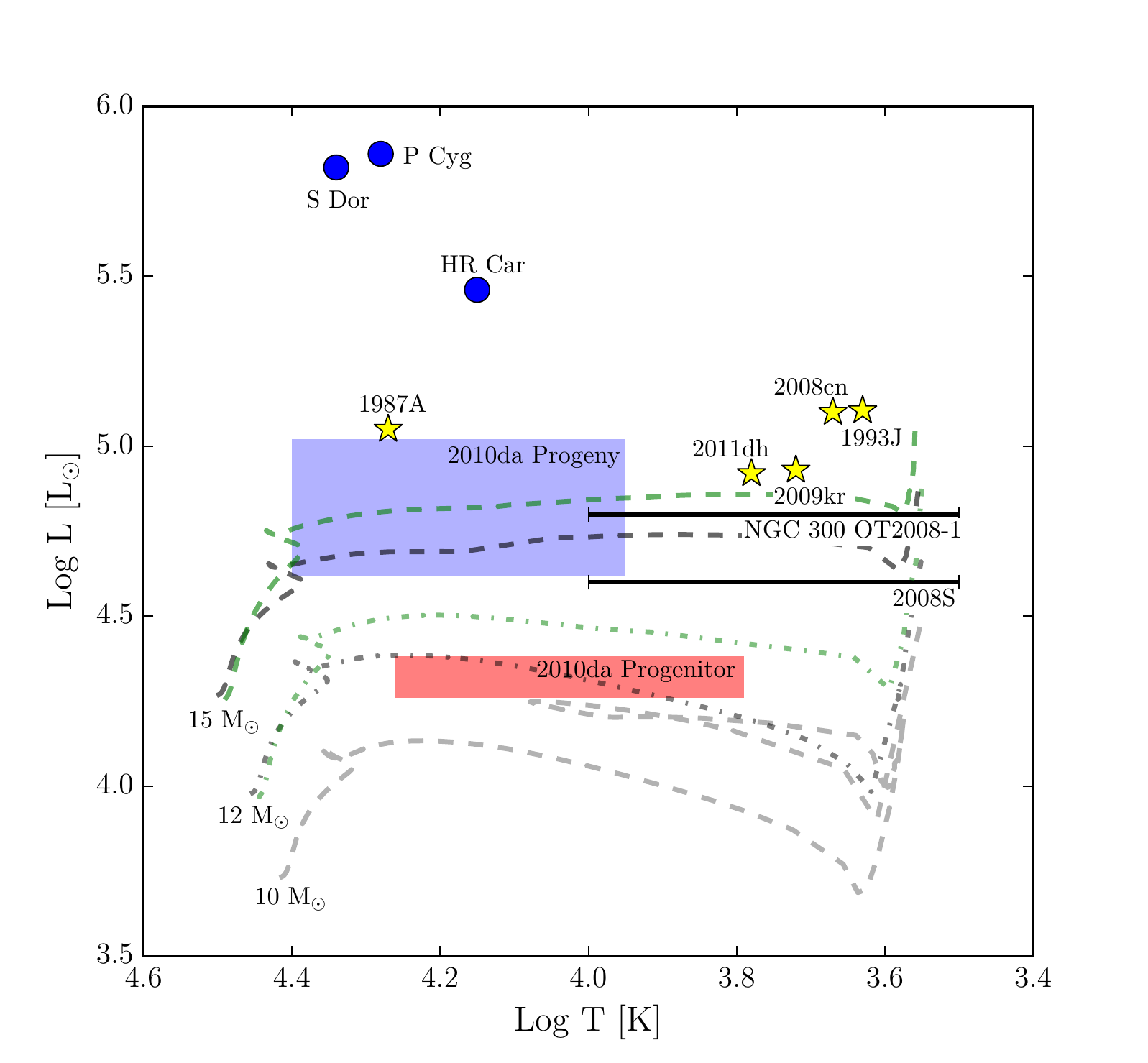}
\caption{Stellar evolutionary tracks of 10 M$_\odot$, 12 M$_\odot$ and 15 M$_\odot$ RSG models with (green) and without (black) rotation from \cite{ekstrom2012} compared to the estimated temperatures and luminosities of progenitor (red) and progeny (blue) of SN 2010da. For comparison, we also show the progenitors of NGC\,300 OT2008-1 \citep{prieto2008ot1}, SN 2008S \citep{prieto2008discovery}, several supernovae and three well-studied LBVs: S Doradus, P Cygni, and HR Carinae \citep{humphreys2011photometric}.}
\label{fig:hr_diagram}
\end{figure*}

To constrain the progeny properties, we use {\tt DUSTY} to model the SEDs around 560 and 1880 days. Again using the constraint from the UVOT light curve, we find that the progeny is hotter than $\approx 8900$ K. Additionally, at temperatures higher than $\approx 25000$ K, the estimated radius becomes atypically small for a supergiant (i.e. $\lesssim 15$ R$_{\odot}$), although we can find acceptable fits beyond this temperature. To reiterate, we have previously ruled out a main sequence star as the progenitor of SN 2010da, meaning that the progeny must also be an evolved supergiant. These temperature and luminosity constraints are shown in an HR diagram in Figure \ref{fig:hr_diagram}. We can again calculate the mass loss rates at these different epochs, this time assuming that a new wind of $v_w\approx 200$ km s$^{-1}$ has formed. We find a mass loss rate of $\dot{M}\approx 3\times10^{-7}$ M$_\odot$ yr$^{-1}$ at 560 days and a slightly larger rate of $\dot{M}\approx 6\times10^{-7}$ M$_\odot$ yr$^{-1}$ at 1880 days. These numbers are consistent with the mass loss rate before the outburst.

The low luminosity, B[e] features and bluer SED are all consistent with a RSG transitioning into a blue loop phase of its evolution \citep{langer1997b}. Additionally, the widths and shapes of the multi-component emission lines are consistent with a newly formed wind interacting with existing mass loss seen in the early stages of a blue-loop phase of a RSG \citep{2008chita}. The blue loop occurs when RSGs evolve off the Hayashi-line towards the BSG regime as their envelope structure shifts from convective to radiative. During this transition, the envelope of the RSG will spin up and the radius will drastically decrease \citep{heger1998}. During this evolution, the star can reach its critical rotation rate and develop a slow equatorial outflow, leading to B[e]-like emission lines.

The environment surrounding the progenitor/progeny is extremely complex, as indicated by the varying estimated electron densities from the Balmer decrement and the existence of forbidden calcium and iron lines. Like many supergiants, SN 2010da might be surrounded by a clumpy wind, which can explain the low- and high-density regions necessary to excite the various emission lines detected in our spectra. The progenitor's dust shell at $\approx$ 10 AU seems to have been at least partially destroyed by the initial transient based on the strong initial UV and X-ray detections. However, the continued infrared excess and SED shape suggests that either some of this dust survived or new dust has since formed at $\approx 6$ AU. The surrounding CSM is carbon rich and irradiated by X-ray/UV emission from the compact binary companion, meaning some dust must be continuously destroyed and formed. During periods of eruptions and enhanced accretion, the UV emission excites coronal iron lines in the CSM, which we observe in the most recent optical spectra taken at 1819 and 1881 days. 

\subsection{SN 2010da as a High Mass X-ray Binary}
\label{sec:hmxb}

Based on the strong X-ray luminosity ($\sim10^{37}$ erg s$^{-1}$) detected well before and after the optical transient, the strong He II 4686\AA\ emission, the coronal iron lines, and the hard X-ray spectrum, we conclude that SN 2010da is in a supergiant X-ray binary system exhibiting B[e] phenomena. A similar conclusion was reached by \cite{binder2011} and \cite{lau2016rising}. However, it is difficult to make a definitive statement about the nature of the compact object itself. The ultraluminous X-ray transient is far above the Eddington limit of a 1.4 M$_\odot$ neutron star, but the hard spectrum ($\Gamma\approx$ 0) and the high X-ray luminosity are consistent with other SGXBs with neutron star companions, such as Vela X-1 \citep{wang2014,binder2011,lewin1997x}. It is possible to explain the super-Eddington luminosity of the initial outburst by invoking beaming along the line of sight or large magnetic fields \citep{mushtukov2015maximum}. In fact, a ULX powered by a neutron star was recently discovered with an X-ray luminosity greater than the peak luminosity of SN 2010da \citep{bachetti2014}. 

SN 2010da also exhibits B[e] phenomena, consistent with a B[e] X-ray binary. Such binaries typically undergo two types of transients: dimmer ($L_X\sim10^{36-37}$ erg s$^{-1}$), shorter ($\tau\sim$ days) Type I outbursts which are associated with the orbital period of the binary, and brighter ($L_X\gtrsim10^{37}$ erg s$^{-1}$), longer ($\tau\gtrsim$ weeks) Type II outbursts which are possibly associated with the disruption of the B[e] disk \citep{reig2011x}. The disk-disruption theory has undergone recent criticism following the discovery of several disks that have remained intact after a Type II outburst \citep{reig2015long}. The duration ($\sim50$ days) and hard spectral index ($\Gamma\sim 0$) of the progeny of SN 2010da are consistent with a Type II outburst \citep{reig2013patterns}. However, the X-ray luminosity during the transient ($L_X\approx 6\times10^{39}$ erg s$^{-1}$) is much more luminous than typical Type II outbursts ($L_X\approx10^{37}-10^{38}$ erg s$^{-1}$). Because little is known about the physical origin of Type II outbursts, we cannot definitely say if SN 2010da is an unusual Type II outburst or a new type of X-ray transient associated with eruptive stellar mass loss.

\subsection{Comparison to Other Dusty ILOTs and Impostors}

Although the canonical model of dusty ILOTs are massive LBVs ejecting dense shells of mass, it has become clear in recent years that these events arise from a variety of progenitors \citep{berger2009,smith2011luminous,kochanek2012unmasking}. Most of the well-studied ILOTs and their progenitors lie in one of two observational classes. The first class is made up of objects with blue and luminous progenitors, such as LBVs or yellow hypergiants (e.g. SN 2009ip, SN 1954J). ILOTs in the ``blue'' class survive their transients and can undergo multiple eruptions. Objects in this class include rare $\eta$ Carinae analogs such as the recent UGC 2773-OT \citep{smith2016persistent} and several ILOTs associated with yellow hypergiants undergoing LBV-like outbursts, like SN Hunt 248 \citep{mauerhan2015sn} and  PSN J09132750+7627410 \citep{tartaglia2016supernova}. The second class of ILOTs is made up of objects with red and extremely cool (T $\sim$ 100s K) progenitor SEDs. These ILOTs appear to be terminal explosions which are potentially electron capture SNe from massive AGB stars (e.g. SN 2008S, NGC\,300 OT2008-1), although other theories exist to explain these events \citep{smith2011luminous,kochanek2012unmasking,2015arXiv151107393A}.

Does the system hosting SN 2010da fit into one of these two classes? We directly compare the progenitor, transient and progeny associated with SN 2010da to two red dusty ILOTs (SN 2008S and NGC\,300 OT2008-1) and two blue ILOTs thought to be LBVs (SN 1954J, or Variable 12 in NGC 2403, and SN 2009ip).

\begin{figure*}[h]
\centering
\includegraphics[width=0.8\textwidth]{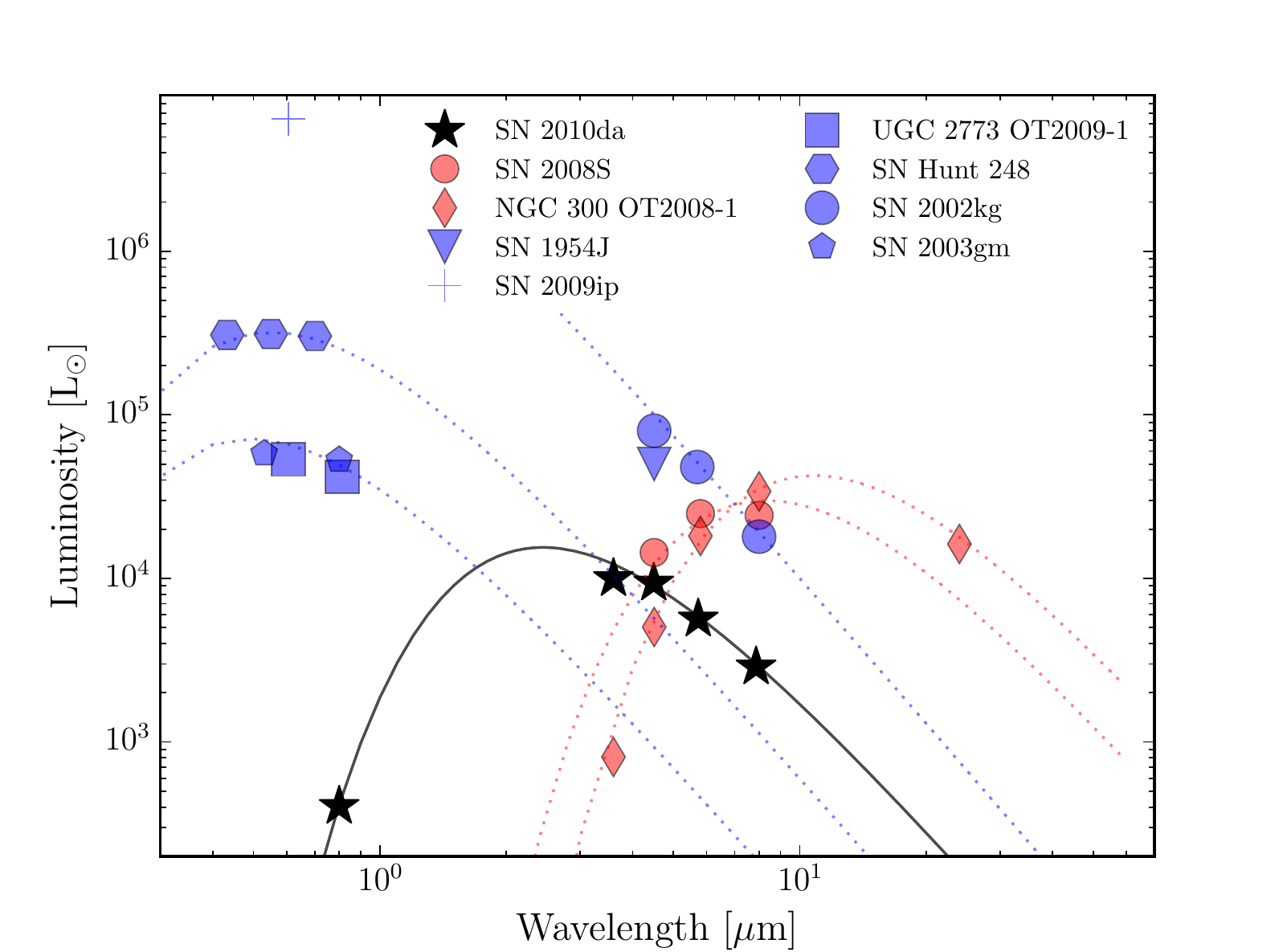}
\caption{Progenitor SEDs of LBV-like (blue) and SN 2008S-like (red) ILOTs compared to the SED of the progenitor of SN 2010da (black). SN 2010da sits between these two classes. The dashed lines are black body fits for select objects to guide the eye.} 
\label{fig:class_seds}
\end{figure*}

SN 2008S and NGC\,300 OT2008-1 had peak absolute magnitudes of $M_V\approx-14$ and $M_V\approx-12$, respectively. These two objects exhibited similar properties and have since faded beyond their initial progenitor luminosities in the IR \citep{2015arXiv151107393A}. SN 2008S and NGC\,300 OT2008-1 had progenitors whose SEDs were consistent with cool circumstellar dust ($T \approx$ 300 $-$ 500 K) and large radii  ($R \approx$ 150 $-$ 350 AU) \citep{prieto2008ot1,khan2010mid}. These temperatures are about four times cooler than the progenitor of SN 2010da ($\approx$ 1500 K), and their estimated radii are about 10 times larger. The luminosities of these progenitors were $\sim 2-3$ times higher than the progenitor of SN 2010da. On the opposite end of the ILOT spectrum lie the blue ILOTs: SN 1954J, a massive star in NGC 2403 which underwent an LBV-like eruption and remains active today, and SN 2009ip, an LBV in NGC 7259 which likely exploded in 2012 \citep{smith2010discovery,margutti2013panchromatic,mauerhan2013unprecedented}. Prior to 1949, the progenitor of SN 1954J had a blue magnitude of M$_b\approx-6.6$ (assuming a distance modulus of 27.6; \citealt{smith2001post}). Similarly, the progenitor of SN 2009ip was extremely bright (M$_\mathrm{Bol}\approx -10$) and variable by as much as one magnitude before its 2009 outburst. Both progenitors of these blue ILOTs are notably brighter and bluer than the progenitor of SN 2010da. The progenitor SED of SN 2010da sits between these two classes, as shown in Figure \ref{fig:class_seds}.

These objects show similar diversity during their transient light curves. Within the first month of discovery, the red ILOTs (SN 2008S and NGC\,300 OT2008-1) experienced a similar decay rate of $\approx$ 0.03 mag d$^{-1}$ \citep{berger2009} --- much more slowly than SN 2010da, which decayed at $\approx 0.1$ mag d$^{-1}$. Although NGC\,300 OT2008-1's light curve steepens at later times (to $\approx$ 0.06 mag d$^{-1}$), it does not exceed the decline rate of SN 2010da. In contrast, the decline rate of SN 2009ip's 2009 outburst within the first month ($\approx 0.2$ mag d$^{-1}$) is faster than that of SN 2010da \citep{smith2010discovery}. In the case of SN 2009ip, such a fast decline rate was attributed by \citealt{smith2010discovery} to the ejection of an optically thick shell, which is not ruled out as a possibility for SN 2010da. 

Spectroscopically, SN 2010da shares features with both the red and blue ILOT classes. For example, the red ILOTs and SN 2010da share similar narrow Balmer and forbidden calcium lines, with H$\alpha$ reaching a maximum width of $\approx$ 1200 km s$^{-1}$. Like NGC\,300 OT2008-1, we detect He I emission in SN 2010da, but we additionally detect He II due to the X-ray/UV-enriched environment from the compact companion. Most notably unlike NGC\,300 OT2008-1, our high resolution spectrum reveals Balmer lines which are weakly asymmetric and lacking any absorption; high-resolution spectra of NGC\,300 OT2008-1 reveal H$\alpha$ emission with clear absorption slightly blueward of rest wavelength \citep{berger2009,bond20092008}. Similarly, the blue ILOTs are also dominated by hydrogen Balmer and Fe II emission (typical of hot LBVs) with FWHM $\approx 550$ km s$^{-1}$ \citep{smith2010discovery, margutti2013panchromatic}.  Unlike SN 2010da, there was no [Ca II] emission detected in SN 2009ip, although [Ca II] emission has been detected in eruptions of cool LBVs such as UGC 2773-OT \citep{smith2010discovery}. Late time spectra of SN 1954J reveal broad H$\alpha$ emission with $\approx 700$ km s$^{-1}$, broader than what is observed in the progeny of SN 2010da.

One of the most notable differences between SN 2010da/the red ILOTs and the blue ILOTs is the fate of their progeny. The blue ILOTs underwent clearly non-terminal eruptions (excluding the 2012 explosion of SN 2009ip; \citealt{margutti2013panchromatic, mauerhan2013unprecedented}). Specifically, recent photometry shows that the progeny of SN 1954J has since faded by $\approx$ 2 mag in the optical and is now consistent with a blackbody with temperature of $\approx$ 6500 K. This has been interpreted as an $\eta$ Car analog which is now shrouded in a dusty nebula similar to $\eta$ Car's Homunculus \citep{smith2001post,van2005supernova}. The most recent SED of SN 1954J is much bluer than that of SN 2010da and suggests a notably higher bolometric luminosity ($\approx 10^5$ L$_\odot$). Based on luminosity and the SED, SN 2010da is unlikely to be an LBV outburst. In contrast, the progenies of SN 2008S and  NGC\,300 OT2008-1 have faded past their progenitors in the IR, leading some authors to argue that they were electron-capture supernovae from super AGB stars \citep{botticella2009sn,thompson2009new,2015arXiv151107393A}. The clear re-brightening of the progeny of SN 2010da several hundred days after the 2010 eruption illustrates that it is not a member of this red class of transients, but its similarities might point to a related progenitor which is entering the last phase of its life. 
 
Thus, SN 2010da is unlike many of the previously studied ILOTs. First, the transient is not energetic enough to be a true LBV outburst. We can roughly estimate the energy radiatively emitted from SN 2010da as $\approx L_{\mathrm{peak}}t_{1.5}$, where $L_{\mathrm{peak}}$ is the peak luminosity and $t_{1.5}$ is the time it takes the transient to dim by 1.5 magnitudes (see \citealt{smith2011luminous}). We estimate $t_{1.5}\lesssim 30$ days based on the upper limit reported by \cite{monard2010}, and we estimate the peak luminosity to be $L_{\mathrm{peak}} = 4.5\times10^{39}$ erg s$^{-1}$. The total radiative energy is thus $\lesssim 10^{46}$ erg. This is less energetic than the typical LBV outburst ($\approx 10^{47}$ erg; \citealt{smith2011luminous}). SN 2010da is also less energetic that the red SN 2008S-like ILOTs, which radiate about $L_{\mathrm{peak}}t_{1.5}\approx 5\times10^{47}$ erg. Additionally, SN 2008S-like events are either terminal or produce progeny that are notably dimmer than their progenitors \citep{2015arXiv151107393A}; the progeny is currently more luminous than its progenitor by a factor of $\approx 5$.

\section{Summary and Conclusions}
We presented comprehensive, multi-wavelength observations (X-ray, UV, optical and IR) of the dusty ILOT SN 2010da, extending thousands of days before and after the outburst. These observations allowed us to study the progenitor, outburst and progeny in great detail. Due to its low luminosity and red SED, SN 2010da seems inconsistent with an LBV outburst as interpreted by \cite{binder2016}. From our dataset, we conclude that SN 2010da was the eruption of a massive star ($\sim 10-12$ M$_\odot$) exhibiting B[e] phenomena. The high-resolution spectra exhibit double-peaked Balmer and Paschen emission lines with narrow components consistent with a pre-existing RSG wind and a newly formed supergiant wind. This suggests that the star responsible for SN 2010da may be a YSG transitioning onto a blue loop. The optical variability and iron/calcium emission indicate a complicated CSM which is repeatedly disturbed by mass loss of the primary star. 

The supergiant responsible for SN 2010da is likely the primary star of a HMXB. The system shows consistently high X-ray luminosity (L$_\mathrm{X}\approx 10^{37}$ erg s$^{-1}$), and during the 2010 event, the system underwent an ultraluminous X-ray outburst ($L_\mathrm{x}\approx 6\times10^{39}$ erg s$^{-1}$). Late time emission of coronal iron lines are fueled by a hot, X-ray- and UV-rich region near this binary. While we cannot make a definitive statement about the nature of the compact object, dedicated and deep X-ray observations may shed light on its nature.

SN 2010da is unique in the heterogeneous class of ILOTs. Its progenitor was dimmer and bluer than the AGB-like progenitors of dusty ILOTs NGC\,300 OT2008-1 and SN 2008S; however, it is notably dimmer and redder than LBVs and yellow hypergiants experiencing similar outbursts. Also unlike other dusty ILOTs and supernova impostors, the progeny of SN 2010da is more luminous than its progenitor in both the IR and optical. The progeny is still undergoing significant outbursts, and continued followup is crucial in understanding the elusive nature of this object. 

Like many ILOTs, SN 2010da marks an important point in stellar evolution of increased activity and mass loss. SN 2010da highlights the diversity of dusty ILOTs and the need for multi-wavelength photometric and high-resolution spectral followup to understand these objects. It is no doubt that future facilities such as LSST will populate the intermediate luminosity gap which currently exists. Extensive spectroscopic followup of current events will allow us to identify archetypes, like SN 2010da, of classes which will arise from these surveys.

\acknowledgments

We thank Rosanne Di Stefano, Sebastian Gomez, Josh Grindlay,  Jeffrey McClintock, and Ramesh Narayan for helpful discussions. We also wish to thank Ian Thompson for obtaining the MIKE spectrum. This paper includes data gathered with the 6.5 meter Magellan Telescopes located at Las Campanas Observatory, Chile. This work is based in part on archival data obtained with the Spitzer Space Telescope, which is operated by the Jet Propulsion Laboratory, California Institute of Technology under a contract with NASA and based on observations obtained at the Gemini Observatory under the Program ID GS-2010A-Q-19, which is operated by the Association of Universities for Research in Astronomy, Inc., under a cooperative agreement with the NSF on behalf of the Gemini partnership: the National Science Foundation (United States), the National Research Council (Canada), CONICYT (Chile), Ministerio de Ciencia, Tecnolog\'{i}a e Innovaci\'{o}n Productiva (Argentina), and Minist\'{e}rio da Ci\^{e}ncia, Tecnologia e Inova\c{c}\~{a}o (Brazil). This work made use of data supplied by the UK Swift Science Data Centre at the University of Leicester. Additionally, the scientific results reported in this article are based in part on data obtained from the Chandra Data Archive. VAV and PKB acknowledge support by the National Science Foundation through a Graduate Research Fellowship.

\bibliographystyle{apj}
\bibliography{vav}


\onecolumn

{\footnotesize
\begin{table}
\centering
\caption{\textit{Spitzer} Photometry}
\begin{tabular}{cccccccc}
Instrument & AOR & PI & Date (UT) & Epoch (Days) & Filter & AB Magnitude \\ \hline\hline
 \textit{IRAC} & 6069760  & Helou  & 2003-Nov-21 &   $-2375$ &      3.6 & 18.77 $\pm$ 0.10 \\
 \textit{IRAC} & 6069760  & Helou & 2003-Nov-21 &   $-2375$ &      4.5 & 18.55 $\pm$ 0.07 \\
 \textit{IRAC} & 6069760  &   Helou  & 2003-Nov-21  & $-2375$ &      5.8 & 19.10 $\pm$ 0.52 \\
 \textit{IRAC} & 6069760 &   Helou    & 2003-Nov-21 & $-2375$ &      8   & 19.51 $\pm$ 0.76 \\
 \textit{MIPS} & 22611456 & Kennicutt & 2007-Jul-16  &   $-1042$ &      24  & $>$ 17.00 \\
 \textit{IRAC} & 22517504 & Kennicutt &2007-Dec-29 &   $ -876$ &      3.6 & 18.79 $\pm$ 0.07 \\
 \textit{IRAC} & 22517504 & Kennicutt & 2007-Dec-29 &   $ -876$ &      4.5 & 18.67 $\pm$ 0.05 \\
 \textit{IRAC} & 22517504 & Kennicutt & 2007-Dec-29 &    $-876$ &      5.8 & 19.50 $\pm$ 0.51 \\
 \textit{IRAC} & 22517504 & Kennicutt & 2007-Dec-29 &    $-876$ &      8   & $>$17.45 \\
 \textit{IRAC} & 31527680 & Freedman & 2009-Dec-21 &    $-153$ &      3.6 & 18.39 $\pm$ 0.07 \\
 \textit{IRAC} & 31527424 & Freedman & 2010-Jan-13 &    $-130$ &      3.6 & 17.84 $\pm$ 0.04 \\
 \textit{IRAC} & 31528448 & Freedman & 2010-Jul-27 &      65 &      3.6 & 17.87 $\pm$ 0.04 \\
 \textit{IRAC} & 31528192 & Freedman & 2010-Aug-16 &      85 &      3.6 & 18.11 $\pm$ 0.05 \\
 \textit{IRAC} & 31527936 & Freedman & 2010-Aug-31 &      100 &      3.6 & 18.36 $\pm$ 0.07 \\
 \textit{IRAC} & 42195968 & Kochanek & 2011-Aug-29 &     463 &      3.6 & 18.68 $\pm$ 0.09 \\
 \textit{IRAC} & 42195968 & Kochanek & 2011-Aug-29 &     463 &      4.5 & 18.85 $\pm$ 0.08 \\
 \textit{IRAC} & 42502912  & Kasliwal & 2012-Jan-14 &     601 &      3.6 & 18.66 $\pm$ 0.08 \\
 \textit{IRAC} & 42195712 & Kochanek& 2012-Aug-10 &     810 &      3.6 & 18.41 $\pm$ 0.07 \\
 \textit{IRAC} & 42195712 & Kochanek& 2012-Aug-10 &     810 &      4.5 & 18.58 $\pm$ 0.07 \\
 \textit{IRAC} & 50572032 & Kasliwal & 2014-Mar-13 &    1390 &      3.6 & 18.41 $\pm$ 0.07 \\
 \textit{IRAC} & 50572032 & Kasliwal & 2014-Mar-13 &    1390 &      4.5 & 18.65 $\pm$ 0.07 \\
 \textit{IRAC} & 50573056 & Kasliwal  & 2014-Sep-05 &    1566 &      3.6 & 18.16 $\pm$ 0.05 \\
 \textit{IRAC} & 50573056 & Kasliwal  & 2014-Sep-05 &    1566 &      4.5 & 18.23 $\pm$ 0.05 \\
 \textit{IRAC} & 50572544  & Kasliwal & 2014-Oct-03 &    1594 &      3.6 & 18.21 $\pm$ 0.06 \\
 \textit{IRAC} & 50572544  & Kasliwal & 2014-Oct-03 &    1594 &      4.5 & 18.28 $\pm$ 0.05 \\
 \textit{IRAC} & 50044672  & Fox & 2014-Oct-14 &    1605 &    3.6 &  18.26   $\pm$ 0.06 \\
 \textit{IRAC} & 50044672 & Fox & 2014-Oct-14 &    1605 &    4.5 &  18.34   $\pm$ 0.04 \\
  \textit{IRAC} & 53022208 & Kochanek & 2015-Feb-09 &    1723 &    3.6 &  18.33   $\pm$ 0.06 \\
  \textit{IRAC} & 52691712 & Kasliwal & 2015-Sep-22&    1948 &    3.6 &  17.91 $\pm$ 0.05 \\
 \textit{IRAC} & 52691712 & Kasliwal & 2015-Sep-22 &    1948 &    4.5 &  18.03 $\pm$ 0.03 \\
  \textit{IRAC} & 52691968  & Kasliwal & 2015-Sep-29 &    1955 &    3.6 &  17.90  $\pm$ 0.04 \\
 \textit{IRAC} & 52691968  & Kasliwal & 2015-Sep-29 &    1955 &    4.5 &  18.03  $\pm$ 0.03 \\
  \textit{IRAC} & 52692224 & Kasliwal  & 2015-Oct-12 &    1968 &    3.6 &  17.89  $\pm$ 0.05 \\
  \textit{IRAC} & 52692224 & Kasliwal  & 2015-Oct-12 &    1968 &    4.5 &  17.99 $\pm$ 0.03 \\
 \textit{IRAC} & 52692480  & Kasliwal & 2016-Feb-22 &    2101 &    3.6 &  18.09  $\pm$ 0.06 \\
  \textit{IRAC} & 52692480  & Kasliwal & 2016-Feb-22 &    2101 &    4.5 &  18.18  $\pm$ 0.04 \\
   \textit{IRAC} & 52692736  & Kasliwal & 2016-Feb-29 &    2108 &    3.6 & 18.13  $\pm$ 0.05 \\
  \textit{IRAC} & 52692736  & Kasliwal & 2016-Feb-29 &    2108 &    4.5 & 18.19  $\pm$ 0.04 \\
   \textit{IRAC} & 52692992  & Kasliwal & 2016-Mar-19 &    2127 &    3.6 &  18.23 $\pm$ 0.06 \\
 \textit{IRAC} & 52692992  & Kasliwal & 2016-Mar-19 &    2127 &    4.5 & 18.22 $\pm$ 0.05 \\
\hline
\label{tab:spitz}
\end{tabular}
\end{table}}

\begin{table}
\centering
\caption{Magellan/\textit{FourStar} Photometry}
\begin{tabular}{cccc}
Date (UT) & Epoch & Filter & AB Magnitude \\ \hline\hline
 2011-Dec-07  & 563 & $J$ &    14.47 $\pm$ 0.09    \\
2011-Dec-07 &  563  & $H$ &    14.17 $\pm$ 0.11     \\
 2011-Dec-07 & 563 & $K_s$ &    14.21 $\pm$ 0.12     \\
2015-Jul-31 &  1895  & $J$ &   14.23 $\pm$ 0.02      \\
 2015-Jul-31 & 1895 & $H$ &  13.76 $\pm$ 0.03       \\
 2015-Jul-31 & 1895 & $K_s$ & 13.57 $\pm$ 0.02   \\
  2015-Aug-18 & 1913 & $H$ & 13.97 $\pm$ 0.01   \\
 2015-Aug-18 & 1913 & $K_s$ &  13.63 $\pm$ 0.02  \\
\hline
\label{tab:fourstar}
\end{tabular}
\end{table}

\begin{table}
\caption{Ground-based Optical Photometry}
\centering
\begin{tabular}{ l c c c c }
    Date (UT) & Epoch (days) & Instrument & Filter & AB Magnitude \\
    \hline\hline
    2008-Sep-09 & $-609$ & IMACS & \textit{i}' & 24.19 $\pm$ 0.20 \\
    2009-Nov-25 & $-179$ & MegaCam & \textit{i}' & $>24.4$ \\
    2009-Nov-25 & $-179$ & MegaCam & \textit{r}' & $>24.4$ \\
    2009-Nov-25 & $-179$ & MegaCam & \textit{g}' & $>24.4$ \\
    2010-Nov-13 & 174 & IMACS & \textit{i}' & 22.97 $\pm$ 0.06 \\
    2010-Nov-13 & 174 & IMACS & \textit{r}' & 22.85 $\pm$ 0.04 \\
    2011-Jan-12 & 234 & LDSS-3 & \textit{i}' & 21.64 $\pm$ 0.03 \\
    2011-Oct-21 & 516 & LDSS-3 & \textit{i}' & 19.77  $\pm$ 0.06 \\
    2011-Oct-21 & 516 & LDSS-3 & \textit{r}' & 19.42  $\pm$ 0.04 \\
    2011-Oct-21 & 516 & LDSS-3 & \textit{g}' & 20.58  $\pm$ 0.03 \\
    2011-Dec-27 & 583 & IMACS & \textit{i}' & 20.29 $\pm$ 0.05 \\
    2011-Dec-27 & 583 & IMACS & \textit{r}' & 20.27 $\pm$ 0.09 \\
    2011-Dec-27 & 583 & IMACS & \textit{g}' & 22.29 $\pm$ 0.20 \\
    2012-May-17 & 725 & LDSS-3 & \textit{i}' & 21.65 $\pm$ 0.04 \\
    2012-May-17 & 725 & LDSS-3 & \textit{r}' & 21.94 $\pm$ 0.04 \\
    2012-May-17 & 725 & LDSS-3 & \textit{g}' & 22.32 $\pm$ 0.03 \\
    2013-Jan-11 & 964 & LDSS-3 & \textit{i}' & 21.70 $\pm$ 0.04 \\
    2013-Jan-11 & 964 & LDSS-3 & \textit{r}' & 20.75 $\pm$ 0.06 \\
    2013-Jul-15 & 1149 & LDSS-3 & \textit{i}' & 20.06 $\pm$ 0.01 \\
    2013-Dec-30 & 1317 & LDSS-3 & \textit{i}' & 20.80 $\pm$ 0.18 \\
    2014-Jun-26 & 1495 & LDSS-3 & \textit{i}' & 21.03 $\pm$ 0.01 \\
    2015-May-15 & 1818 & IMACS & \textit{i}' & 20.03 $\pm$ 0.02 \\
    2015-May-15 & 1818 & IMACS & \textit{r}' & 18.89 $\pm$ 0.13 \\
    2015-Jul-17 & 1881 & IMACS & \textit{i}' & 20.18 $\pm$ 0.02 \\
    2015-Jul-17 & 1881 & IMACS & \textit{r}' & 19.87 $\pm$ 0.02 \\
    2015-Jul-17 & 1881 & IMACS & \textit{g}' & 20.99 $\pm$ 0.03 \\
    2015-Aug-01 & 1896 & IMACS & \textit{i}' & 19.71 $\pm$ 0.05 \\
    2015-Aug-01 & 1896 & IMACS & \textit{r}' & 20.09 $\pm$ 0.05 \\
    2015-Aug-01 & 1896 & IMACS & \textit{g}' & 20.74 $\pm$ 0.08 \\
    \hline  
    \label{tab:ldss}
\end{tabular}
\end{table}

\begin{table}[t]
\caption{\textit{HST} Photometry}
\centering
\begin{adjustbox}{width=0.7\textwidth}
\begin{tabular}{c c c c c c}
Start Date (UT) & Epoch & Proposal ID & PI & Filter & AB Magnitude\\
\hline\hline
  2012-Jul-18 & 787 & 12450 & Kochanek & F814W & 20.63 $\pm$ 0.03\\
  2014-Jul-02 & 1501 & 13515 & Binder & F606W & 20.68 $\pm$ 0.02\\
  2014-Jul-02 & 1501 & 13515 & Binder & F814W & 20.99 $\pm$ 0.03\\
\hline  
\label{tab:hst}
\end{tabular}
\end{adjustbox}
\end{table}

\begin{table}[t]
\caption{Ground-based Optical Spectroscopy}
\centering
\begin{adjustbox}{width=0.7\textwidth}
\begin{tabular}{lccccc}
\hline\hline
Date (UT) & Epoch (days) & Instrument & Exposure (s) & Grating/Grism & Resolution (\AA)  \\
\hline
2010-May-25 & 2 & GMOS-S & 1200 & R400 & 4 \\
2010-May-25 & 2 & GMOS-S & 600 & B600 & 4 \\
2010-Jun-07 & 15 & MIKE & 1800 & R2 & 0.3 \\
2010-Jun-07 & 15 & MIKE & 1800 & R2.4 & 0.4 \\
2010-Jul-02 & 40 & GMOS-S & 1200 & R400 & 4 \\
2010-Jul-02 & 40 & GMOS-S & 900 & B600 & 4 \\
2011-Oct-21 & 516 & MagE & 1200 & - & 2 \\
2011-Dec-29 & 585 & IMACS & 300 & 300-17.5 & 5 \\
2015-May-16 & 1819 & IMACS & 1500 & 300-17.5 & 5 \\
2015-Jul-17 & 1881 & IMACS & 1800 & 300-17.5 & 5 \\
    \hline  
    \label{tab:spec}
\end{tabular}
\end{adjustbox}
\end{table}

\begin{table}[t]
\caption{\textit{Swift UVOT}/ Photometry}
\centering
\begin{adjustbox}{width=0.9\textwidth}
\begin{tabular}{cccccccc}
\hline\hline
Date & Epoch & \multicolumn{6}{c}{ AB Magnitude }\\
(UT)&  & UVW2 & UVM2 & UVW1 & U & B & V \\
\hline
2010-May-26 & 3 & 18.27 $\pm$ 0.08 & 18.06 $\pm$  0.07 & 17.48 $\pm$  0.07 & 16.72 $\pm$  0.07 & 16.48 $\pm$  0.07 & 16.23 $\pm$  0.07 \\ 
2010-May-27 & 4 & 18.20 $\pm$ 0.08 & 17.96 $\pm$  0.07 & 17.43 $\pm$  0.071 & 16.67 $\pm$  0.07 & 16.43 $\pm$  0.07 & 16.17 $\pm$  0.08 \\ 
2010-May-28 & 5 & 18.33 $\pm$ 0.08 & 18.10 $\pm$  0.07 & 17.57 $\pm$  0.07 & 16.73 $\pm$  0.067 & 16.53 $\pm$  0.07 & 16.31 $\pm$  0.08 \\ 
2010-May-31 & 8 & 18.89 $\pm$ 0.09 & 18.58 $\pm$  0.08 & 17.97 $\pm$  0.076 & 17.21 $\pm$  0.07 & 16.92 $\pm$  0.08 & 16.61 $\pm$  0.09 \\ 
2010-Jun-01 & 9 & 19.31 $\pm$ 0.10 & 18.79 $\pm$  0.08 & 18.25 $\pm$  0.08 & 17.46 $\pm$  0.08 & 17.20 $\pm$  0.08 & 17.00 $\pm$  0.11 \\ 
2010-Jun-03 & 11 & 19.43 $\pm$ 0.11 & 18.93 $\pm$  0.09 & 18.41 $\pm$  0.10 & 17.48 $\pm$  0.08 & 17.45 $\pm$  0.10 & 17.05 $\pm$  0.11 \\ 
2010-Jun-05 & 14 & 19.61 $\pm$ 0.11 & 19.30 $\pm$  0.09 & 18.61 $\pm$  0.085 & 17.77 $\pm$  0.08 & 17.63 $\pm$  0.10 & 17.35 $\pm$  0.12 \\ 
2010-Jun-07 & 16 & 19.89 $\pm$ 0.12 & 19.58 $\pm$  0.10 & 18.81 $\pm$  0.092 & 18.06 $\pm$  0.09 & 17.93 $\pm$  0.11 & 17.75 $\pm$  0.16 \\ 
2010-Jun-09 & 18 & 20.07 $\pm$ 0.13 & 19.62 $\pm$  0.10 & 1.46 $\pm$  0.0 & 18.24 $\pm$  0.10 & 18.14 $\pm$  0.13 & 17.78 $\pm$  0.16 \\ 
2010-Jun-11 & 20 & 20.13 $\pm$ 0.14 & 19.83 $\pm$  0.11 & 19.11 $\pm$  0.10 & 18.53 $\pm$  0.11 & 18.56 $\pm$  0.17 & 18.02 $\pm$  0.20 \\ 
2010-Jun-16 & 24 & 20.49 $\pm$ 0.16 & 20.18 $\pm$  0.13 & 19.46 $\pm$  0.123 & 18.71 $\pm$  0.12 & 18.87 $\pm$  0.21 & 18.48 $\pm$  0.27 \\ 
2010-Jun-17 & 25 & 20.53 $\pm$ 0.17 & 20.10 $\pm$  0.15 & 19.61 $\pm$  0.15 & 18.92 $\pm$  0.14 & 18.85 $\pm$  0.20 & 18.59 $\pm$  0.32 \\ 
2010-Jun-18 & 26 & 20.51 $\pm$ 0.16 & 20.15 $\pm$  0.13 & 19.52 $\pm$  0.12 & 19.08 $\pm$  0.14 & 19.03 $\pm$  0.23 &  $>$ 19.01 \\ 
2010-Jun-21 & 29 & 21.05 $\pm$ 0.26 & 20.57 $\pm$  0.24 & 19.89 $\pm$  0.148 & 19.02 $\pm$  0.14 &  $>$ 19.71 &  $>$ 18.25 \\ 
2010-Jun-26 & 34 & 20.8 $\pm$ 0.24 & 20.08 $\pm$  0.22 & 19.84 $\pm$  0.181 & 19.28 $\pm$  0.20 &  $>$ 19.42 &  $>$ 17.98 \\ 
2010-Jul-02 & 40 & 21.21 $\pm$ 0.25 & 21.13 $\pm$  0.23 & 20.54 $\pm$  0.219 & 19.76 $\pm$  0.21 &  $>$ 19.82 &  $>$ 19.00 \\ 
2011-Oct-26 & 521 & $>$ 21.46 &  $>$ 20.86 &  $>$ 20.72 & 20.84 $\pm$  0.30 &  $>$ 19.43 &  $>$ 18.62 \\ 
2011-Oct-27 & 522 & $>$ 22.32 &  $>$ 21.02 &  $>$ 20.84 &  $>$ 20.30 &  $>$ 19.53 &  $>$ 18.72 \\ 
2011-Oct-28 & 523 & $>$ 21.64 & 21.85 $\pm$  0.33 &  $>$ 20.92 &  $>$ 20.39 &  $>$ 19.60 &  $>$ 18.80 \\ 
\hline

\hline  
    \label{tab:uvotphot}
\end{tabular}
\end{adjustbox}
\end{table}

\begin{table}[t]
\caption{\textit{Chandra} Photometry}
\centering
\begin{adjustbox}{width=0.7\textwidth}
\begin{tabular}{c c c c c c c}
Start Date & Epoch & Proposal ID & PI & Counts & Detection Significance & 0.3 - 10 keV Flux \\
(UT) &  & &  &  & ($10^{-15}$ erg s$^{-1}$ cm$^{-2}$)\\
\hline\hline
  2010-Sep-24 & 123 & 12238 & Williams & 77 $\pm$ 9 & 21$\sigma$ & $50.7^{+8.9}_{-9.2}$\\
  2014-May-16 & 1453 & 16028 & Binder & 7 $\pm$ 3 & 3$\sigma$ & $5.6^{+2.4}_{-3.3}$\\
  2014-Nov-17 & 1638 & 16029 & Binder & 140 $\pm$ 12 & 52$\sigma$ & $142^{+12}_{-28}$\\
\hline  
\label{tab:chandra}
\end{tabular}
\end{adjustbox}
\end{table}


\begin{table}
\caption{Summary of SN 2010da Blackbody Fits}
\centering
\begin{tabular}{c c c c }
	\hline
     & Progenitor & Epoch 1 & Epoch 9 \\
    \hline
    $T_C$ & 1500 $\pm$ 40 K & 3230 $\pm$ 490 K & 2760 $\pm$ 250 K \\
   $T_H$ & \ldots & 9440 $\pm$ 280 K & 9080 $\pm$ 330 K \\
    $R_C$ & 9.4 $\pm$ 0.5 AU & 9.5 $\pm$ 2.9 AU & 10.5 $\pm$ 1.6 AU \\
   $R_H$ & \ldots & 1.59 $\pm$ 0.14 AU & 1.25 $\pm$ 0.13 AU \\
\hline
    \label{tab:BB_fits}
\end{tabular}
\end{table}

{\footnotesize
\begin{landscape}
\renewcommand{\arraystretch}{0.6}
\begin{longtabu}{@{\extracolsep{\fill}} X[c]X[c]X[c]X[c]X[c]X[c]X[c]}
\caption{Catalog of Spectral Lines Identified in Spectra of SN 2010da}\\
Line & Epoch & Profile & Line Center  & EW  & FWHM  & Flux \\
 &  &  &(\AA) &(\AA) & (km s\textsuperscript{-1}) & (10\textsuperscript{-16}\hspace{.1cm}erg\hspace{.1cm}s\textsuperscript{-1}\hspace{.1cm}cm\textsuperscript{-2})\\
\hline\hline
\endfirsthead

\multicolumn{7}{r}{Continued on previous page}\\
Line & Epoch & Profile & Line Center  & EW  & FWHM  & Flux \\
 &  &  &\AA &\AA & km s\textsuperscript{-1} & 10\textsuperscript{-16}\hspace{.1cm}erg\hspace{.1cm}s\textsuperscript{-1}\hspace{.1cm}cm\textsuperscript{-2}\\
\hline\hline
\endhead

\hline \multicolumn{7}{r}{{Continued on next page}} \\ 
\endfoot

\multicolumn{7}{l}{{Note: Profiles are described as Gaussian (G), Lortenzian (L), Double Gaussian (DG), Gausian and Lorentzian blend (LG) or P-Cygni profile (P-Cyg). }}\\
\multicolumn{7}{l}{{Lines with missing epochs are not covered in the spectral range, while lines with ND are within the spectral range but not detected. Numbers in parentheses are 1$\sigma$ errors. }}\\
\multicolumn{7}{l}{{The reported line center and FWHM errors do not account for uncertainties due to instrument resolution.}}
\endlastfoot

Fe I 3752.88 & 2 & SP & 3752.4 & $-$0.2 (0.1) & $\lesssim$ 400 &  7 (1)\\
Fe I 3774.76 & 2 & SP & 3772.1 & $-$1.30 (0.02) & 420 (40) & 6.46 (0.08) \\
Fe I 3800.61 & 2 & DP & 3802.0, & $-$2.21 (0.09) & $<$ 400 & 10.56 (0.09) \\
 &  &  & 3804.0 & & 450 &  \\
 & 15 & SP & 3800.1 & $-4 (1)$  & 420 (10) & \ldots \\
  & 40 & SP & 3800.0 & $-1.4 (0.3)$ & $\lesssim$ 300 & 0.5 (0.3) \\
H$\eta$ & 2 & SP & 3839.0 & $-1.83 (0.08)$ & 570 (60) & 7 (2) \\
 & 15 & SP & 3838.2 & $-$1.8 (0.8) & 440 (20) & \ldots \\
 & 40 & SP & 3836.9 & $-$2.4 (0.2) & 140 (50) & 1.0 (0.4) \\
H$\zeta$ & 2 & P-Cyg & 3885.3, & \ldots & \ldots & \ldots\\
 &  & & 3891.9 &  &  & \\
 & 15 & DP & 3890.1 & $-$9 (3) & 430 (10) & \ldots\\
  &  &  & 3891.8 & & 65 (9) & \\
 & 40 & SP & 3891.2 & $-$3.1 (0.2) & 350 (40) & 2.6 (0.2)\\
Fe I 3935.81 & 2 & P-Cyg & 3932.9, & \ldots & \ldots & \ldots \\
\& Ca II K  &  &  & 3937.9 &  &  &  \\
 & 15 & SP & 3935.77 & $-$5 (1) & 390 (10) & \ldots \\
 & 40 & SP & 3936.97 & $-$15 (2) & 470 (30) & 1.1 (0.3) \\
H$\epsilon$ & 2 & DP & 3976.0 & $-$3.8 (0.1) & 1100 (300) & 27 (1)\\
\& Ca II H &  &  & 3973.13 &  & 490 (40) & \\
 & 15 & DP & 3971.8 & $-$10 (2) & 490 (20) & \ldots\\
  &  & & 3972.8 &  & 130 (20) & \\
 & 40 & SP & 3971.36 & $-$18 (1) & $<$ 300 & 5.2 (0.3)\\
H$\delta$ & 2 & SP & 4104.72 & $-$5.52 (0.07) & 560 (20) & 32 (1)\\
 & 15 & DP & 4102.9 & $-$6 (1) & 42 (2) & \ldots\\
  &  &  & 4104.1 &  & 449 (8) & \ldots\\
 & 40 & SP & 4103.76 & $-$12 (1) & 480 (30) & 2.2 (0.5)\\
  & 516 & SP & 4014.11 & $-$12 (4) & 110 (10) & \ldots\\
Fe II 4174.62 & 2 & ND & \ldots & \ldots & \ldots & \ldots \\
 & 15 & SP & 4174.86 & $-$0.23 (0.05) & 100 (10) & \ldots \\
 & 40 & SP & 4174.81 & $-$5.2 (0.2) & 220 (30) & 2.5 (0.3) \\
 &516 & SP & 4175.29 & $-$5.0 (0.7) & 150 (20) & \ldots \\
Fe II 4180.03 & 2 & ND & \ldots & \ldots & \ldots & \ldots \\
 & 15 & SP & 4180.15 & $-$0.7 (0.1) & 220 (30) & \ldots \\
 & 40 & SP & 4179.66 & $-$15 (3) & 590 (70) & 1.3 (0.3) \\
 & 516 & SP & 4180.15 & $-$6 (2) & 900 (200) & \ldots \\
Fe I 4237.12 & 2 & SP & 4236.24 & $-$0.75 (0.02) & 500 (50) & 4.3 (0.7) \\
 & 15 & SP & 4234.9 & $-$2.2 (0.2) & 109 (6) & \ldots \\
 & 40 & SP &4234.6 & $-$7.8 (0.6) & 350 (50) & 1.6 (0.2) \\
  & 516 & SP & 4235.0 & $-$8 (1) & 450 (60) & \ldots \\
H$\gamma$ & 2 & SP & 4342.9 & $-$8.8 (0.1) & 590 (20) & 53 (1)\\
    & 15 & DP & 4342.2 & $-$11 (2) & 458 (9) & \ldots\\
        &  &  & 4342.8 &  & 56 (2) & \\
 & 40 & SP & 4342.0 & $-$29 (4) & 180 (10) & 8.6 (0.2)\\
 & 516 & DP & 4341.4 & $-$19 (3) & 510 (200) & \ldots \\
  &  &  & 4342.9 & & 120 (10) &  \\
    & 1819 & SP & 4343.2 & $-$9 (3) & 400 (60) & 2.2 (0.3)\\
    & 1881 & SP & 4343.2 & $-$20 (20) & 630 (140) & 7 (1)\\
{[O III]} 4363& 2 & ND & \ldots & \ldots & \ldots & \ldots \\
\& {[Fe IX]} 4359 & 15 & SP & 4364.3 & $-$0.027 (0.008) & 40 (4) & \ldots \\
 \& Fe II & 40 & ND & \ldots & \ldots & \ldots & \ldots \\
 & 516 & ND & \ldots & \ldots & \ldots & \ldots \\
 & 1819 & SP & 4365.8 & $-$5 (1) & 260 (50) & 2.2 (0.1) \\
 & 1881 & SP & 4365.9 & $-$30 (20) & 500 (100) & 7.2 (0.2) \\
Fe I 4384.77 & 2 & P-Cyg & 4382.0 & \ldots & \ldots & \ldots \\
 &  &  & 4388.3 &  &  &  \\
 & 15 & SP & 4387.1 & $-$0.8 (0.1) & 280 (20) & \ldots \\
 & 40 & SP & 4387.0 & $-$2.8 (0.2) & 460 (80) & 1.4 (0.1) \\
  & 516 & SP & 4388.0 & $-$3.3 (0.4) & 360 (40) & \ldots \\
 & 1819 & ND & \ldots & \ldots & \ldots & \ldots \\
Fe I  4416.36 & 2 & SP & 4418.6 & $-$0.27 (0.01) & 340 (80) & 3.1 (0.5) \\
 & 15 & SP & 4418.5 & $-$2.0 (0.2) & 280 (10) & \ldots \\
 & 40 & SP & 4417.3 & $-$3.1 (0.3) & 450 (70) & 0.8 (0.3) \\
  & 516 & SP & 4418.0 & $-$3.8 (0.4) & 390 (40) & \ldots \\
 & 1819 & ND & \ldots & \ldots & \ldots & \ldots \\
  & 1881 & ND & \ldots & \ldots & \ldots & \ldots \\
 He I 4471.5 & 2 & P-Cyg & 4467.0 & \ldots & \ldots & \ldots\\
   & 2 &  & 4474.5 &  &  & \\
   & 15 & SP & 4473.4 & $-$2.7 (0.3) & 370 (10) & \ldots\\
  & 40 & SP & 4473.3 & $-$3.7 (0.3) & 340 (50) & 0.8 (0.1) \\
& 516 & SP & 4474.6 & $-$2.9 (0.6) & 250 (40) & \ldots \\
  & 1819 & ND & \ldots & \ldots & \ldots & \ldots \\
  & 1881 & ND & \ldots & \ldots & \ldots & \ldots \\
Fe I 4585 & 2 & SP & 4586.0 & $-$1.30 (0.03) & 340 (80) & 6 (1) \\
 & 15 & SP & 4586.0 & $-$7 (1) & 330 (50) & \ldots \\
  & 40 & SP & 4585.9 & $-$2.4 (0.7) & 440 (50) & 1.6 (0.7) \\
  & 516 & SP & 4585.9 & $-$1.4 (0.3) & 130 (20) & \ldots \\
  & 1819 & SP & 4584.8 & $>$-$6$ & $\lesssim$ 230 & $<0.9$ \\
  & 1881 & SP & 4587.8 & $>$-$6$ & 600 (100) & $<1.0$ \\
He II 4686 & 2 & SP & 4683.0 & $-$3.2 (0.6) & 550 (30) & 23 (1) \\
 & 15 & SP & 4687.2 & $-$3.0 (0.2) & 270 (10) & \ldots \\
 & 40 & SP & 4686.7 & $-$7.5 (0.5) & 330 (20) & 2.7 (0.2) \\
 & 516 & SP & 4687.6 & $-$7 (2) & 270 (20) & \ldots \\
 & 1819 & SP & 4687.5 & $-$12 (4) & 480 (60) & 1.6 (0.3) \\
 & 1881 & SP & 4687.6 & $-$30 (10) & 560 (70) & 4.5 (0.7) \\
H$\beta$ & 2 & SP & 4863.6 & $-$20.9 (0.4) & 530 (20) & 160.8 (0.7)\\
 & 15 & DP & 4862.9 & $-$16 (1) & 66 (3) & \ldots\\
  &  &  & 4864.6 &  & 499 (7) & \\
 & 40 & SP & 4863.4 & $-$110 (10) & 360 (10) & 36.3 (0.2)\\
 & 516 & DP & 4863.7 & $-$70 (10) & 106 (4) & \ldots\\
  &  &  & 4861.7 &  & 540 (60) & \ldots\\
   & 1819 & SP & 4863.8 & $-$80 (20) & 450 (10) &  12.5 (0.2)\\
   & 1881 & SP & 4864.0  & $-$170 (4) & 620 (10) &  24.7 (0.3)\\
{[O III]} 4959 & 2 & ND & \ldots & \ldots & \ldots & \ldots \\
   & 15 & SP & 4960.3 & $>-0.4$ & 50 (10) & \ldots \\
  & 40 & SP & 4960.0 & $-$2.3 (0.1) & 280 (30) & 0.4 (0.1) \\
  & 516 & ND & \ldots & \ldots & \ldots & \ldots \\
 & 1819 & SP & 4960.9 & $-$8 (1) & 300 (40) & 1.8 (0.3) \\
 & 1881 & SP & 4958.0 & $-$20 (7) & 530 (80) & 3.1 (0.3) \\
{[O III]} 5007 & 2 & SP & 5007.9 & $-$0.22 (0.03) & $<$ 240 & 2.1 (0.5) \\
 & 15 & SP & 5008.0 & $-$1.4 (0.1) & 63 (2) & \ldots \\
 & 40 & SP & 5008.0 & $-$4.4 (0.5) & $\lesssim$ 240 & 1.7 (0.3) \\
  & 516 & ND & \ldots & \ldots & \ldots & \ldots \\
 & 1819 & SP & 5009.1 & $-$25 (7) & 350 (20) & 5.6 (0.3) \\
 & 1881 & DP & 5008.3 & $-$80 (10) & $\lesssim$ 300 & 9.4 (0.3) \\
  &  &  & 5012.3 &  & 500 (20) &  \\
Fe I  5017.87 & 2 & SP & 5020.15 & $-$2.36 (0.02) & 440 (20) & 17.6 (0.2) \\
 & 15 & DP & 5021.1 & $-$5.1 (0.8) & 430 (20) & 2.9 (0.2) \\
  &  &  & 5020.0 &  & 37 (3) & \\
 & 40 & SP & 5019.8 & $-$7 (1) & 350 (50) & 3.2 (0.1) \\
  & 516 & DP & 5019.7 & $-$7 (1) & 420 (50) & \ldots \\
    &  &  & 5020.7 &  & $\lesssim$ 90 &  \\
  & 1819 & ND & \ldots & \ldots & \ldots & \ldots \\
    & 1881 & ND & \ldots & \ldots & \ldots & \ldots \\
{[Fe VII]} 5159 & 2 & SP & 5159.8 & $>0.07$ & 170 (50) & 1.1 (0.1) \\
\& Fe II & 15 & SP & 5160.2 & $-$1.2 (0.4) & $<80$ & \ldots \\
 & 40 & ND & \ldots & \ldots & \ldots & \ldots\\
  & 516 & SP & 5160.6 & $<0.6$ & 130 (30) & \ldots\\
 & 1819 & SP & 5161.0 & $-$6 (1) & $<$ 290 & 0.4 (0.1) \\
  & 1881 & SP & 5161.0 & $-$5 (2) & 400 (100) & 1.0 (0.2) \\
 {[Fe VII]} 5276 & 2 & $-$ & \ldots & \ldots & \ldots & Note: on chip \\
    & 15 & SP & 5277.4 & $-$0.4 (0.1) & 300 (30) & \ldots \\
    & 40 & $-$ & \ldots & \ldots & \ldots & Note: on chip \\
    & 516 & SP & 5276.6 & $-$2.9 (0.4) & 60 (10) & \ldots \\
   & 1819 & SP & 5277.7 & $>-4$ & $<400$ & $<0.7$ \\
   & 1881 & SP & 5274.0 & $>-4$ & $<600$ & $<2$ \\
{[O I]} 5577 & 2 & ND & \ldots & \ldots & \ldots & \ldots \\
 & 15 & ND & \ldots & \ldots & \ldots & \ldots \\
 & 40 & ND & \ldots & \ldots & \ldots & \ldots \\
  & 516 & SP & 5576.5 & $-$0.08 (0.04) & 60 (20) & \ldots \\
 & 1819 & SP & 5577.2 & $-$11 (2) & 180 (20) & 0.20 (0.08) \\
 & 1881 & SP & 5577.3 & +4 (1) & 300 (50) & $-$1.0 (0.1) \\
{[N II]} 5755 & 2 & ND & \ldots & \ldots & \ldots & \ldots \\
 \& Fe II & 15 & SP & 5756.3 & $-$1.2 (0.1) & 200 (100) & 0.4 (0.1) \\
 & 40 & ND & \ldots & \ldots & \ldots & \ldots \\
  & 516 & ND & \ldots & \ldots & \ldots & \ldots \\
 & 1819 & SP & 5756.5 & $-$0.9 (0.5) & 200 (100) & 0.29 (0.03) \\
 & 1881 & SP & 5753.6 & $-$6 (2) & 800 (200) & 1.5 (0.4) \\
He I 5877& 2 & SP & 5890.0 & $-$5.1 (0.1) & 450 (20) & 36.6 (0.7) \\
 & 15 & DP (LG) & 5877.3 & $-$6.2 (0.8) & 30 (4) & \ldots \\
  & &  & 5878.7 & & 370 (10) &  \\
 & 40 & SP & 5878.2 & $-$7.6 (0.7) & 290 (20) & 6.7 (0.3) \\
  & 516 & SP & 5877.5 & $-$8 (2) & 170 (20) & \ldots \\
   & 1819 & SP & 5878.0 & $-$11 (9) & 360 (80) & 0.4 (0.3) \\
   & 1881 & SP & 5878.5 & $-$32 (7) & 500 (40) & 4.3 (0.5) \\
{[Fe VII]} 6086 & 2 & SP & 6088.4 & $-$0.07 (0.01) & 260 (60) & 0.2 (0.1) \\
 & 15 & SP & 6088.3 & $>-0.3$ & 240 (30) & \ldots \\
 & 40 & SP & 6087.6 & 0.83 (0.03) & 230 (30) & 0.60 (0.06) \\
  & 516 & ND & \ldots & \ldots & \ldots & \ldots \\
 & 1819 & SP & 6090.0 & $-$3.3 (0.8) & 280 (40) & 0.7 (0.2) \\
 & 1881 & SP & 6088.8 & $-$13 (4) & 400 (50) & 1.81 (0.3) \\
? & 2 & SP (L) & 6280.5 & +0.3 (0.1) & 280 (50) & 1.4 (0.4) \\
 & 15 & ND & \ldots & \ldots & \ldots & \ldots \\
 & 40 & ND & \ldots & \ldots & \ldots & \ldots \\
 & 516 & ND & \ldots & \ldots & \ldots & \ldots \\
 & 1819 & ND & \ldots & \ldots & \ldots & \ldots \\
 & 1881 & ND & \ldots & \ldots & \ldots & \ldots \\
{[O I]} 6300 & 2 & ND & \ldots & \ldots & \ldots & \ldots \\
  & 15 & SP & 6302.1 & $-$0.53 (0.02) & 270 (20) & \ldots \\
  & 40 & SP & 6303.8 & $-$0.73 (0.05) & 240 (60) & 0.28 (0.05) \\
  & 516 & SP & 6302.5 & $-$1.4 (0.1) & 150 (20) & \ldots \\
  & 1819 & SP & 6302.2 & $-$7.6 (0.7) & 260 (50) & 0.84 (0.04) \\
  & 1881 & SP & 6303.2 & $-$9 (2) & 400 (70) & 1.0 (0.3) \\
{[Fe X]} 6374 & 2 & SP & 6380.8 & $-$0.7 (0.1) & 700 (100) & 1.2 (0.2) \\
   & 15 & SP & 6375.5 & $-$0.8 (0.1) & 230 (20) & \ldots \\
   & 40 & P-Cyg & 6360.4 & \ldots & \ldots & \ldots \\
      &  &  & 6374.9 &  &  &  \\
  & 516 & ND & \ldots & \ldots & \ldots & \ldots \\
  & 1819 & SP & 6377.3 & $-$18 (3) & 270 (20) & 2.7 (0.1) \\
 & 1881 & SP & 6377.8 & $-$26 (4) & 430 (40) & 3.9 (0.2) \\
H$\alpha$ & 2 & SP & 6566.9 & $-$105 (3) & 556 (6) & 660 (1)\\
  & 15 & TP & 6564.9,  & $-$302 (2) & 69.6 (0.4), & \ldots \\
    & &  & 6565.0, &  & 494 (3), &  \\
      &  &  & 6568.4 & & 1074 (4) &  \\
  & 40 & SP (L) & 6567.7 & $-$680 (30) & 490 (10) & 292.7 (0.2) \\
  & 516 & DP & 6566.7 & $-$260 (50) & 140 (4), & \ldots\\
    &  &  & 6570.4 & & 280 (20) & \\
  & 585 & SP & 6566.7 & $-$150 (30) & 470 (20) & 9.5 (0.3) \\
   & 1819 & SP & 6566.4 & $-$980 (50) & 360 (8) & 164.0 (0.3) \\
   & 1881 & DP & 0 6563.3,  & $-$1300 (100) & 1700 (200),  & 256.7 (0.5) \\
      &  & & 0 6565.8 &  & 570 (10) &  \\
He I 6678& 2 & SP & 6684.0 & $-$3.98 (0.02) & 580 (10) & 17.4(0.9) \\
 & 15 & SP & 6681.2 & $-$4.5 (0.2) & 350 (10) & \ldots \\
 & 40 & SP & 6682.7 & $-$5.9 (0.2) & 440 (20) & 1.5 (0.2) \\
  & 516 & SP & 6681.3 & $-$7.2 (0.7) & 210 (10) & \ldots \\
   & 1819 & SP & 6681.0 & $-$5 (1) & 260 (50) & 0.8 (0.1)\\
      & 1881 & SP & 6680.3 & $-$1.2 (0.7) & 440 (90) & 1.3 (0.2) \\
He I 7065& 2 & SP & 7070.5 & $-$5.44 (0.05) & 500 (10) & 30.4 (0.4) \\
 & 15 & DP (LG) & 7067.2, & $-$5.4 (0.8) & 28 (3), & \ldots \\
  &  & & 7068.4 & & 40 (20) &  \\
 & 40 & SP & 7070.0 & $-$9.8 (0.2) & 428 (9) & 4.46 (0.07) \\
  & 516 & SP & 7069.0 & $-$6.6 (0.5) & 200 (10) & \ldots \\
  & 1819 & SP & 7067.7 & $-$10 (1) & 250 (20) & 0.8 (0.1)\\
    & 1881 & SP & 7068.1 & $-$18 (2) & 280 (20) & 1.5 (0.3)\\
He I 7281& 2 & ND & \ldots & \ldots & \ldots & \ldots \\
 & 15 & SP & 7284.8 & $-$1.0 (0.1) & 270 (40) & \ldots \\
 & 40 & ND & \ldots & \ldots & \ldots & \ldots \\
  & 516 & ND & \ldots & \ldots & \ldots & \ldots \\
  & 1819 & ND & \ldots & \ldots & \ldots & \ldots\\
    & 1881 & ND & \ldots & \ldots & \ldots & \ldots\\
{[CaII]} 7291& 2 & ND & \ldots & \ldots & \ldots & \ldots \\
  & 15 & SP & 7293.7 & $>-0.2$ & $\lesssim$ 16 & \ldots \\
 & 40 & ND & \ldots & \ldots & \ldots & \ldots \\
 & 516 & SP & 7293.3 & $-$0.5 (0.1) & $\lesssim$ 40 & \ldots \\
 & 1819 & ND & \ldots & \ldots & \ldots & \ldots \\
  & 1881 & ND & \ldots & \ldots & \ldots & \ldots \\
{[CaII]} 7323 & 2 & ND & \ldots & \ldots & \ldots & \ldots \\
  & 15 & SP & 7326.1 & $-$0.123 (0.003) & $\lesssim$ 20 & \ldots \\
 & 40 & ND & \ldots & \ldots & \ldots & \ldots \\
 & 516 & SP & 7326.0 & $-$0.9 (0.1) & $\lesssim$ 60 & \ldots \\
 & 1819 & ND & \ldots & \ldots & \ldots & \ldots \\
  & 1881 & ND & \ldots & \ldots & \ldots & \ldots \\
O I 7774& 2 & P-Cyg & 7765.4, & \ldots & \ldots & \ldots \\
 &  & & 7783.1 &  &  &  \\
 & 15 & P-Cyg & 7769.8, & \ldots & \ldots & \ldots \\
  &  & & 7780.4 & &  & \\
 & 40 & P-Cyg & 7771.7,  & \ldots & \ldots & \ldots \\
  &  &  & 7781.0 &  &  &  \\
  & 516 & SP & 7779.4 & $-$1.9 (0.2) & 90 (30) & \ldots \\
  & 1819 & ND & \ldots & \ldots & \ldots & \ldots \\
    & 1881 & ND & \ldots & \ldots & \ldots & \ldots \\
{[Fe XI]} 7892 & 2 & ND & \ldots & \ldots & \ldots & \ldots \\
  & 15 & ND & \ldots & \ldots & \ldots & \ldots \\
 & 40 & ND & \ldots & \ldots & \ldots & \ldots \\
  & 516 & ND & \ldots & \ldots & \ldots & \ldots \\
 & 1819 & SP & 7895.3 & $-$6 (2) & 230 (40) & 1.0 (0.2) \\
 & 1881 & SP & 7895.0 & $-$17 (2) & 400 (40) & 1.9 (0.5) \\
O I 8446 & 2 & SP (L) & 8451.4 & $-$5.6 (0.2) & 160 (10) & 6 (1) \\
 & 15 & DP & 8449.1, & $-$10.3 (0.9) & 35 (2), & \ldots \\
  & &  & 8451.3 & & 470 (20) &  \\
 & 40 & SP (L) & 8452.1 & $-$30 (10) & 190 (20) & 1.3 (0.2) \\
  & 516 & SP & 8450.0 & $-$7 (2) & 270 (20) & \ldots \\
    & 585 & SP & 8450.3 & $-$7 (1) & 270 (60) & 0.4 (0.1) \\
  & 1819 & SP & 8449.8 & $-$6 (1) & 270 (60) & 6.3 (0.7) \\
    & 1881 & SP & 8449.4 & $-$24 (4) & 190 (10) & 6.5 (0.8) \\
Ca II & 2 & P-Cyg & 8495.6,  & \ldots & \ldots & \ldots \\
\& Pa 8500 &  &  & 8509.1 &  &  &  \\
 & 15 & DP & 8500.6, & $-$11.1 (0.6) & 36 (1), & \ldots \\
  &  &  & 8503.7 &  & 418 (7) &  \\
 & 40 & SP & 8503.6 & $-$26 (1) & 346 (9) & 9.0 (0.1) \\
  & 516 & DP & 8504.3, & $-$16 (1) & 60 (5), & \ldots \\
    &  &  & 8500.8 &  & 30 (2) &  \\
    & 585 & SP & 8502.3 & $-$9 (3) & 180 (30) & 1.0 (0.2) \\
  & 1819 & SP & 8501.4 & $-$7 (1) & 620 (140) & 1.4 (0.1) \\
    & 1881 & SP & 8498.5 & $-$12 (2) & 510 (60) & 2.03 (0.09) \\
Ca II & 2 & SP & 8549.0 & $-$2.30 (0.06) & 420 (20) & 19.0 (0.9) \\
\& Pa 8544  & 15 & DP & 8544.7, & $-$9 (2) & 39 (1),  & \ldots \\
&  &  & 8547.7 & & 520 (70) &  \\
 & 40 & SP & 8547.7 & $-$20 (2) & 320 (8) & 7.56 (0.05) \\
   & 516 & DP & 8545.6,  & $-$16 (1) & 43 (3),  & \ldots \\
      &  & & 8546.6 &  & 270 (20) &  \\
   & 585 & SP & 8547.7 & $-$7 (3) & 240 (50) & 0.7 (0.1) \\
  & 1819 & SP & 8547.3 & $-$7 (1) & 190 (40) & 0.6 (0.1) \\
      & 1881 & SP & 8545.7 & $-$7 (1) & 250 (40) & 1.21 (0.06) \\
Ca II  & 2 & SP $-$ flat & 8669.9 & $-$4.8 (0.1) & 540 (20) & 22.6 (0.5) \\
\& Pa 8664 & 15 & DP & 8664.9, & $-$9.3 (0.9) & 37 (4),  & \ldots \\
&  & & 8668.0 &  & 460 (40) &  \\
 & 40 & SP & 8668.2 & $-$16.7 (0.8) & 301 (9) & 5.30 (0.08) \\
   & 516 & DP & 8665.5,  & $-$18 (3) & 58 (4),  & \ldots \\
      &  &  & 8668.2 &  & 340 (30) &  \\
    & 585 & SP & 8667.0 & $-$8 (2) & 190 (30) & $<0.4$ \\
  & 1819 & SP & 8666.0 & $-$7 (2) & 130 (40) & 1.3 (0.3) \\
      & 1881 & DP & 8666.7, & $-$11 (1) & 380 (70),  & 1.4 (0.1) \\
    &  &  & 8672.2 & & 400 (300) & \\
Pa 8753 & 2 & SP & 8755.0 & $-$1.88 (0.06) & 410 (60) & 5 (1) \\
    & 15 & DP & 8753.1,  & $-$2.7 (0.2) & 32 (5),  & \ldots\\
 &  &  & 8755.9 &  & 160 (10) & \\
    & 40 & SP & 8760.0 & $-$13.1 (0.8) & 550 (20) & 3.6 (0.1)\\
    & 516 & SP & 8754.8 & $-$3.7 (0.3) & 150 (20) & \ldots\\
    & 1819 & ND & \ldots & \ldots & \ldots & \ldots\\
    & 1881 & SP & 8755.0 & $-$6 (2) & 180 (60) & 1.0 (0.2)\\
Pa 8865 & 2 & P-Cyg & 8851.9, & \ldots & \ldots & \ldots \\
 &  &  & 8871.7 &  &  &  \\
    & 15 & DP & 8865.4,  & $-$4.7 (0.6) & 100 (10),  & \ldots\\
    &  &  & 8867.8 &  & 400 (30) & \\
    & 40 & SP & 8868.0 & $-$6 (1) & 360 (50) & 3.0 (0.3)\\
    & 516 & SP & 8867.4 & $-$4.0 (0.5) & 100 (10) & \ldots\\
    & 1819 & SP & 8866.5 & $-$1.1 (0.9) & 100 (50) & 2.0 (0.3)\\
    & 1881 & SP & 8865.6 & $-$7.2 (0.9) & 600 (100) & 1.3 (0.2)\\
Pa 9017 & 2 & SP & 9024.3 & $-$6.5 (0.5) & 840 (70) & 26.7 (0.9) \\
    & 15 & SP & 9020.5 & $-$4.9 (0.5) & 410 (50) & \ldots\\
    & 40 & SP & 9021.6 & $-$11 (1) & 430 (30) & 2.3 (0.1)\\
    & 516 & SP & 9019.0 & $-$4.6 (0.5) & 180 (20) & \ldots\\
    & 1819 & ND & \ldots & \ldots & \ldots & \ldots\\
    & 1881 & SP & 9018.3 & $-$7 (1) & 290 (50) & 0.3 (0.1)\\
Pa 9232 & 2 & SP & 9234.1 & $-$2.4 (0.2) & 820 (240) & 8 (1) \\
    & 15 & DP & 9231.7, & $-$2.4 (0.2) & 49 (3), & \ldots\\
     &  &  & 9235.2 & & 110 (10) & \\
    & 40 & DP & 9227.1, & $-$7 (1) & 180 (20), & 2.8 (0.2)\\
  &  &  & 9236.3 & & 630 (80) & \\
    & 516 & SP & 9233.0 & $-$5.0 (0.7) & 150 (20) & \ldots\\
    & 1819 & SP & 9233.0 & $-$8 (1) & 150 (30) & 0.40 (0.09)\\
    & 1881 & SP & 9232.1 & $-$5 (1) & 290 (50) & 0.7 (0.3)\\
Pa$\epsilon$ 9548 & 2 & SP & 9557.1 & $-$0.77 (0.02) & 180 (40) & 50 (1)\\
    & 40 & SP & 9553.1 & $-$19 (4) & 500 (60) & 5.4 (0.2)\\
    & 1881 & SP & 9548.8 & $-$9 (2) & 280 (60) & 4.2 (0.5)\\
Pa$\delta$ 10052 & 2 & SP & 10057.1 & $-$7.2 (0.2) & 152 (7) & 32 (2) \\
    & 40 & SP & 10051.7 & $-$0.7 (0.9) & 430 (70) & 3 (1)\\
    & 1881 & SP & 10052.6 & $-$10 (3) & 230 (40) &  3.7 (0.5)\\
    \hline  
    \label{tab:specdetails}
\end{longtabu}
\end{landscape}}

\end{document}